\documentclass[10pt,twocolumn,letterpaper]{article}

\usepackage{cvpr}              
\usepackage[pagebackref,breaklinks,colorlinks,allcolors=cvprblue]{hyperref}
\usepackage{multirow}
\definecolor{cvprblue}{rgb}{0.21,0.49,0.74}
\usepackage[pagebackref,breaklinks,colorlinks,allcolors=cvprblue]{hyperref}

\def\paperID{} 
\def\confName{CVPR}
\def\confYear{2026}

\title{TIACam: Text-Anchored Invariant Feature Learning with Auto-Augmentation for Camera-Robust Zero-Watermarking}

\author{
Abdullah All Tanvir \quad
Agnibh Dasgupta \quad
Xin Zhong \\
Department of Computer Science \\
University of Nebraska Omaha \\
{\tt\small atanvir@unomaha.edu \quad adasgupta@unomaha.edu \quad xzhong@unomaha.edu}
}

\usepackage{comment}

\usepackage{xcolor}

\newcommand\mn{{TIACam}\xspace}

\begin{document}
\maketitle
\begin{abstract}
Camera recapture introduces complex optical degradations, such as perspective warping, illumination shifts, and Moiré interference, that remain challenging for deep watermarking systems. We present TIACam, a text-anchored invariant feature learning framework with auto-augmentation for camera-robust zero-watermarking. The method integrates three key innovations: (1) a \textit{learnable auto-augmentor} that discovers camera-like distortions through differentiable geometric, photometric, and Moiré operators; (2) a \textit{text-anchored invariant feature learner} that enforces semantic consistency via cross-modal adversarial alignment between image and text; and (3) a \textit{zero-watermarking head} that binds binary messages in the invariant feature space without modifying image pixels. This unified formulation jointly optimizes invariance, semantic alignment, and watermark recoverability. Extensive experiments on both synthetic and real-world camera captures demonstrate that TIACam achieves state-of-the-art feature stability and watermark extraction accuracy, establishing a principled bridge between multimodal invariance learning and physically robust zero-watermarking.
\end{abstract}

\vspace{-0.5em}
    
\vspace{-0.5em}
\section{Introduction}
\label{sec:intro}
\vspace{-0.5em}

Image watermarking is a covert technique for embedding hidden watermark information into digital images to ensure copyright protection, content authentication, and ownership verification. 
Traditional watermarking methods modify the image either in the spatial or transform domain, balancing imperceptibility: the degree to which the watermark remains visually invisible, and robustness: the ability to correctly extract the watermark after image distortions. 
In contrast, zero-watermarking eliminates direct image modification by associating a watermark with intrinsic features extracted from the original image~\cite{fierro2019robust}. 
This paradigm preserves visual imperceptibility while enabling reliable watermark verification or extraction through feature matching. 
Recent advances in deep learning have greatly expanded the scope of both embedding-based and zero-watermarking schemes, where neural networks automatically learn discriminative and robust features that support watermark extraction under diverse transformations.

\begin{figure}[t]
\centering
\includegraphics[width=1.0\columnwidth]{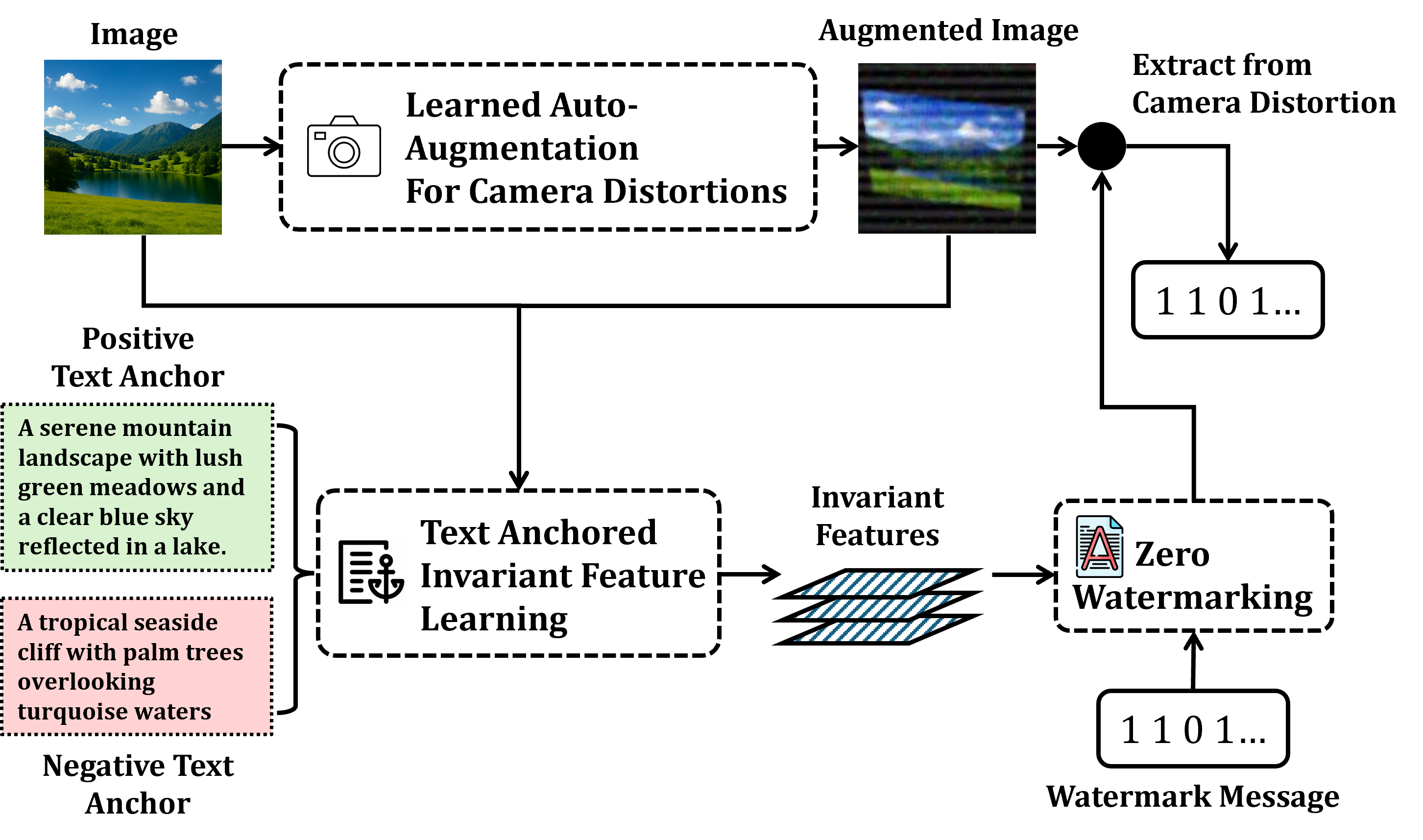}
\vspace{-1.5em}
\caption{Concept of the proposed \mn.}
\label{fig:concept}
\vspace{-2.0em}
\end{figure}

Among the applications, watermark extraction from camera-captured images remains particularly challenging. 
Unlike synthetic distortions such as rotation or blur, camera recapture introduces compound and spatially coupled degradations such as perspective warping, illumination variation, sensor noise, and color imbalance (see Fig.~\ref{fig:concept}). 
To address these effects, recent learning-based watermarking methods~\cite{tancik2020stegastamp, fang2022pimog} train with a fixed ``camera noise'' layer that approximates physical distortions. 
This strategy improves camera robustness by augmenting manually simulated distortions directly into the training process. 
In parallel, another line of robust watermarking research leverages pretrained visual feature extractors, such as those from self-supervised (SSL) contrastive learning frameworks~\cite{caron2021emerging}, to achieve robustness without direct noise modeling. 
These approaches draw inspiration from the broader goal of representation learning, capturing high-level semantics invariant to pixel variations.

However, existing methods face several limitations. 
First, manually designing camera noise layers is inherently restrictive: real-world optical distortions are environment-dependent, non-linear, and contextually coupled, making them difficult to simulate through fixed augmentations. 
Second, although pretrained SSL models provide some representations, they are not explicitly optimized for watermarking; feature robustness emerges as a side effect rather than a targeted objective. 
Finally, even with these advances, watermark extraction accuracy under real camera capture remains a major unresolved challenge. 

To this end, we propose \textbf{\mn} (\textbf{T}ext-Anchored \textbf{I}nvariant learning with \textbf{A}uto-augmentation for \textbf{Cam}era robustness), a unified framework that learns camera-robust invariant features for zero-watermarking. 
Our approach introduces three key contributions: 
(1) a \emph{learnable auto-augmentor} that automatically discovers realistic camera-like distortions through differentiable noise modules; 
(2) a \emph{text-anchored invariant feature learner} that enforces semantic stability across distortions via cross-modal adversarial training; and 
(3) a \emph{zero-watermarking head} is adopted to bind binary messages to the invariant feature space, achieving high watermark recovery accuracy under synthetic and real-world camera captures. 
By jointly learning invariance and semantic alignment, \mn advances both the theoretical understanding and the practical robustness of camera-based zero-watermarking.

\vspace{-0.5em}
\section{Related Work}
\label{sec:related_work}
\vspace{-0.5em}

Deep learning–based image watermarking has primarily focused on enhancing robustness. 
For example, 
HiDDeN~\cite{zhu2018hidden} introduced an end-to-end encoder–decoder framework that learns invisible perturbations to embed data in images, achieving robustness against distortions such as blur, cropping, and JPEG compression. 
MBRS~\cite{jia2021mbrs} enhanced JPEG robustness by mixing real and simulated compression during training and integrating squeeze-and-excitation and diffusion modules within an encoder–decoder framework. 
DWSF~\cite{guo2023practical} proposed a dispersed embedding and synchronization–fusion strategy for high-resolution and arbitrary-size images, improving robustness against geometric and combined attacks. 
MuST~\cite{wang2024must} developed a multi-source tracing watermarking scheme that detects and resynchronizes multiple embedded watermarks to trace composite images back to their original sources. 
WOFA~\cite{liu2025watermarking} designed a hierarchical embedding–extraction model with a comprehensive distortion layer to withstand partial image theft, enabling watermark recovery from arbitrary image fragments. 
Watermark extraction from camera-captured images has been particularly challenging due to complex, coupled degradations arising from sensor noise, illumination variation, and perspective misalignment~\cite{tancik2020stegastamp, fang2022pimog}. 
To address this, 
proposals such as StegaStamp~\cite{tancik2020stegastamp} and PIMoG~\cite{fang2022pimog} incorporate manually designed camera-like noise layers into the encoder-decoder training process, simulating projection, blur, color shifts, and compression to mimic real capture conditions. 
While these approaches improve camera robustness through manual noise augmentation, their performance remains constrained by the limited diversity and realism of handcrafted distortion models, which struggle to generalize across heterogeneous capture environments.

Recent studies on deep robust watermarking and zero-watermarking have shifted toward using robust features that remain stable under distortions rather than directly embedding watermarks in pixel space. 
A common strategy is to leverage pre-trained networks as fixed feature extractors, assuming that their learned representations are inherently robust to perturbations. 
Some works~\cite{fierro2019robust,he2023shrinkage,han2024application,li2024zwnet} adopt convolutional or transformer-based backbones to extract semantic features from the host image and fuse them with encoded watermark patterns through simple correlation schemes. 
More recent approaches~\cite{vukotic2020classification,fernandez2022watermarking} extend this idea by exploiting self-supervised models such as DINO~\cite{caron2021emerging} to embed or verify watermarks in pretrained feature spaces, gaining additional robustness to common distortions. 
However, feature stability arises passively from pre-training objectives rather than being explicitly optimized for watermarking robustness.

To move beyond reliance on pre-trained representations, recent works have started to explicitly train invariant features for watermarking. 
One line of research introduces text-guided invariance~\cite{ahtesham2025text}, where cosine similarity between image and caption embeddings is minimized to maintain semantic consistency across visual transformations. 
Another, exemplified by InvZW~\cite{tanvir2025invzw}, formulates zero-watermarking as a distortion-adversarial task in which a discriminator distinguishes clean from distorted image features. 
In contrast, the proposed \mn establishes a novel unified adversarial dynamic that jointly learns camera-like distortions, enforces cross-modal semantic alignment, and achieves robustness surpassing both text-only and distortion-only invariance paradigms.

\vspace{-0.5em}
\section{\mn}
\label{sec:methodology}
\vspace{-0.5em}

\begin{figure*}[ht]
\vspace{-1.0em}
\centering
\includegraphics[width=1.9\columnwidth]{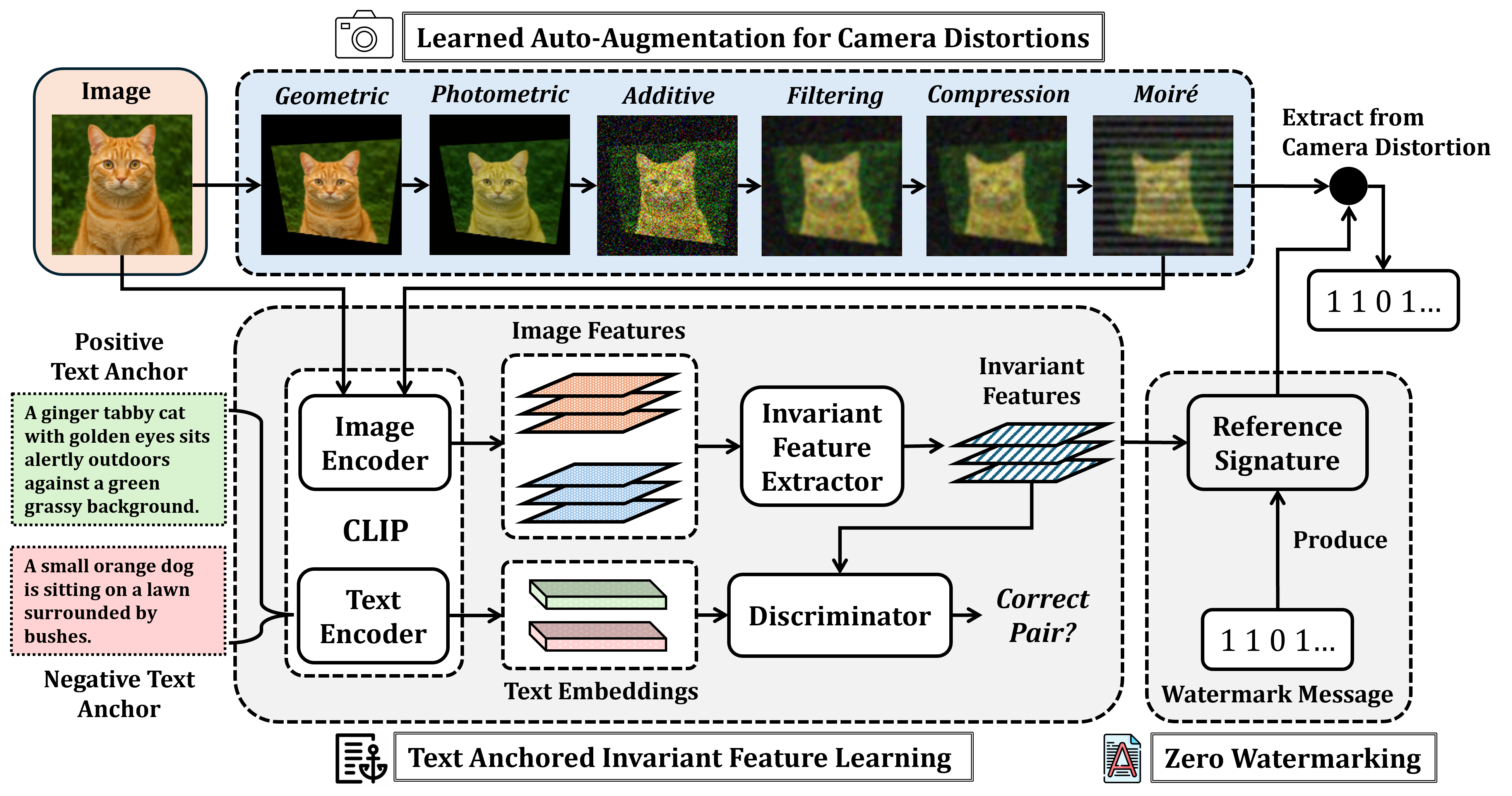}
\vspace{-0.5em}
\caption{
Overview of the proposed \mn. 
Given an input image $x$ and its positive anchor caption $T$ with a negative caption $\tilde{T}$, a distorted image $\hat{x} = \mathcal{T}_{\text{aug}}(x)$ is generated using the learned auto-augmentor $\mathcal{T}_{\text{aug}}(\cdot)$. 
All inputs are encoded by the CLIP encoders to obtain 768-D features, which are refined by the invariant feature extractor $f_{\theta}(\cdot)$ into 1024-D invariant representations. 
Paired samples $(f_{\theta}(x), g_{\tau}(T))$, $(f_{\theta}(\hat{x}), g_{\tau}(T))$, and $(f_{\theta}(\hat{x}), g_{\tau}(\tilde{T}))$ are used to train a discriminator $D_{\psi}(\cdot)$ that distinguishes real from fake associations, while $f_{\theta}$ is adversarially optimized both against $D_{\psi}$ for semantic alignment and against $\mathcal{T}_{\text{aug}}$ for robustness. 
For zero-watermarking, the invariant feature $f_{\theta}(x)$ is projected onto reference codes $C$, and watermark bits are predicted as $\hat{W} = \sigma(f_{\theta}(x)^{\top} C)$ for reliable extraction.
}
\label{fig:proposed_method}
\vspace{-1.5em}
\end{figure*}

As shown in Fig.~\ref{fig:proposed_method}, the proposed framework comprises three modules, an Auto-Augmentor, a Text-Anchored Invariant Feature Learner, and a Zero-Watermarking Head, working jointly in a loop. 
The Auto-Augmentor generates realistic camera degradations through differentiable camera noise operators, producing learnable and augmented image variants. 
The Text-Anchored Invariant Feature Learner aligns both the original and augmented images with their textual descriptions via a CLIP backbone and a lightweight invariant feature extractor–discriminator pair. Its training combines adversarial objectives against the discriminator (for semantic alignment) and against the Auto-Augmentor (for robustness). 
The learned invariant features, stable under camera-induced perturbations, are then used by the Zero-Watermarking Head to bind binary watermark messages to latent feature signatures, enabling reliable extraction without modifying image pixels.

\vspace{-0.5em}
\subsection{Learned Auto-Augmentation}
\vspace{-0.5em}

To simulate realistic distortions encountered in camera-captured or physically altered images, we design a fully differentiable Auto-Augmentor composed of learnable modules for geometric, photometric, additive, filtering, compression, and Moir\'e noise transformations (see Fig.~\ref{fig:proposed_method}). 
Each module is implemented as a parameterized neural operator, enabling gradients to flow through the entire augmentation pipeline during adversarial training. 
Unlike traditional fixed augmentations, the proposed Auto-Augmentor dynamically learns distortion distributions that most challenge feature invariance, thereby modeling realistic camera variations. 

\vspace{-1.5em}
\paragraph{Geometric Module.}
This module applies spatial transformations to capture camera motion, rotation, scaling, and projection effects. 
Following the differentiable perspective transformation layer from~\cite{khatri2022perspective}, we define
\(
x' = \mathcal{T}_{\text{geo}}(x; \theta_{\text{geo}}), \quad 
\mathcal{T}_{\text{geo}}(x)(u,v) = x(A [u,v,1]^\top),
\)
where $A \in \mathbb{R}^{3\times3}$ is a learnable perspective matrix. 
This formulation allows continuous updates of parameters such as shearing, stretching, or viewpoint shifts through gradient descent.

\vspace{-1.5em}
\paragraph{Photometric Module.}
This module models illumination changes via differentiable brightness, contrast, and gamma transformations:
\(
\mathcal{T}_{\text{photo}}(x'') = \alpha \cdot x''^\gamma + \beta,
\)
where $\alpha$, $\gamma$, and $\beta$ (scalar or per-channel) are learnable parameters representing contrast, gamma, and brightness. 
Bounded optimization of these parameters ensures physically plausible color transformations.

\vspace{-1.5em}
\paragraph{Additive Noise Module.}
Sensor-like degradation is introduced through differentiable additive noise:
\(
\mathcal{T}_{\text{noise}}(x''') = x''' + \sigma \cdot z,\quad 
z \sim \mathcal{N}(0,1),\ \sigma>0,
\)
where the reparameterization trick maintains gradient flow through the stochastic operation. 
For discrete artifacts (e.g., salt-and-pepper), a Gumbel-softmax approximation~\cite{jang2017categorical} is employed to ensure differentiability.

\vspace{-1.5em}
\paragraph{Filtering Module.}
simulate optical blur and lens smearing, a learnable convolutional kernel $K$ performs 
\(
x'' = \mathcal{T}_{\text{filter}}(x') = K * x',
\)
where $K$ represents Gaussian or motion blur kernels constrained by a normalization loss for numerical stability. 
Gradients propagate through $K$, enabling the network to learn distortion kernels that maximally reduce feature consistency.

\vspace{-1.5em}
\paragraph{Compression Module.}
To approximate lossy compression artifacts such as quantization and blocking, we implement a differentiable surrogate of JPEG compression~\cite{zhu2018hidden}. 
Given input $x$, the discrete cosine transform $\mathcal{D}(x)$ is followed by a smooth quantization function~\cite{semenov2025smooth}:
\(
\mathcal{Q}(z) = \lfloor z \rfloor + \sigma(\alpha(z-\lfloor z \rfloor)) - 0.5,
\)
where $\sigma(\cdot)$ is the sigmoid function and $\alpha$ controls smoothness. 
A trainable frequency-domain mask $M \in [0,1]^{H\times W}$ further modulates $\mathcal{D}(x)$ as 
$\mathcal{D}'(x)=M\odot\mathcal{D}(x)$, 
allowing the model to learn adaptive compression patterns that reflect real-world bandwidth or codec distortions.

\vspace{-1.5em}
\paragraph{Moiré Distortion Module.}
To emulate the interference patterns that arise from sensor–display misalignment, we introduce a differentiable Moiré generator:
\(
\mathcal{T}_{\text{moire}}(x''') = x''' + \alpha \cdot \sin\!\big(2\pi(f_x u + f_y v) + \phi\big),
\)
where $(u,v)$ are pixel coordinates, $(f_x, f_y)$ denote spatial frequencies sampled from a learnable range, and $\phi$ is a random phase offset. 
The amplitude $\alpha$ is a trainable parameter that controls pattern strength, while $(f_x, f_y)$ are differentiable with respect to the augmentation loss.
This formulation enables gradient-based optimization of pattern frequency and intensity, allowing the augmenter to reproduce realistic periodic interference effects observed in camera recapture.

\vspace{-1.5em}
\paragraph{Modular Composition.}
The full augmentation function is defined as a sequential composition of six differentiable modules:
\(
\mathcal{T}_{\text{aug}} =
\mathcal{T}_{\text{comp}} \circ
\mathcal{T}_{\text{filter}} \circ
\mathcal{T}_{\text{add}} \circ
\mathcal{T}_{\text{photo}} \circ
\mathcal{T}_{\text{geo}} \circ
\mathcal{T}_{\text{moire}},
\quad
\hat{x} = \mathcal{T}_{\text{aug}}(x; \Theta),
\)
where $\Theta = \{\theta_{\text{geo}}, \alpha, \beta, \gamma, \sigma, M, \alpha_{\text{moire}}\}$ denotes all learnable parameters across modules. 
During training, $\mathcal{T}_{\text{aug}}$ is optimized adversarially to generate distortions that most disrupt feature alignment, while the invariant feature extractor $f_{\theta}$ is trained to preserve semantic consistency. 
This adversarial interplay enables the model to discover \emph{which distortions matter} and \emph{how to remain stable under them}, capturing the complex diversity of real camera and environmental perturbations.



\vspace{-0.5em}
\subsection{Text-Anchored Invariant Feature Learning}
\label{subsec: adversarial_training}
\vspace{-0.5em}

Modeling camera distortions is inherently difficult because real-world degradations are compound - perspective shifts, compression, lighting changes, and sensor noise often co-occur in unpredictable ways. 
In the feature space, rather than modeling every possible distortion explicitly, we approach the problem from the ground up: learning what remains stable, the semantic meaning of the image.  
If a watermark is embedded not in pixel values but in the meaning of the image, it will remain intact as long as the meaning is preserved, a property naturally satisfied under most camera distortions. 
This motivates our central feature proposal: a model that can match multimodality media (in our case, images and texts) robustly across distortions has, in a sense, learned to ``understand'' the content. 
Hence, the watermark can be embedded within this invariant feature space that captures the meaning of the image rather than in the raw image domain.

\vspace{-1.5em}
\paragraph{Proposed Semantic Invariance Principle.}
We define a composite feature extractor $f_\theta$ that consists of a frozen CLIP image encoder and a trainable invariant feature extractor built on top of it. 
Given an input image $I$ and its text anchor $T$, the image and text embeddings are obtained as $F=f_\theta(I)$ and $E=g_\tau(T)$, where $g_\tau$ is the CLIP text encoder. 
For a content-preserving transformation $\mathcal{T}$ (e.g., camera recapture, compression, illumination shift), the extractor is trained to satisfy
\(
F \approx f_\theta(\mathcal{T}(I)),
\)
ensuring that both the original and distorted images share a consistent, semantics-preserving representation. 
The text embedding $E$ serves as a stable anchor that captures the core meaning while discarding transient visual details. 
Because caption-like text anchors are abstract yet content-specific, they provide a high-level invariant axis for aligning visual features across distortions.

This formulation follows the Information Bottleneck (IB) principle~\cite{tishby2015deep}, which seeks representations that preserve task-relevant semantics while discarding nuisance variations. 
Here, the visual feature $F$ serves as a bottleneck between the raw image and its semantic anchor $E$, optimized as
\(
\max I(F; E) - \beta I(F; I),
\)
where $I(\cdot;\cdot)$ denotes mutual information. 
Maximizing $I(F;E)$ enforces semantic consistency with the text description, while penalizing $I(F;I)$ reduces sensitivity to low-level appearance changes. 
This encourages $F$ to encode the image’s invariant meaning rather than its unstable visual realization.

\vspace{-1.5em}
\paragraph{Adversarial Cross-Modal Alignment Training.}
We implement the semantic–invariance principle as a cross–modal min–max game that discriminates between \emph{matched} and \emph{mismatched} image–text pairs. 
Given a batch $\mathcal{B}=\{(I,T^+)\}$ with each image $I$ and its positive text anchor $T^+$, we sample a semantically related but incorrect caption $T^-$ as a negative text anchor. 
A distorted image view $I' = A_{\phi}(I)$ is generated by the Auto-Augmentor. 
Embeddings are obtained as
\(
F = f_{\theta}(I),\quad 
F' = f_{\theta}(I'),\quad 
E^+ = g_{\tau}(T^+),\quad 
E^- = g_{\tau}(T^-),
\)
where $f_{\theta}$ denotes the composite image feature extractor (CLIP encoder plus invariant feature extractor) and $g_{\tau}$ is the frozen CLIP text encoder.

A pair discriminator $D_{\psi}$ receives an image–text feature pair and predicts whether they convey the same semantic content (\emph{real}) or not (\emph{fake}).
This formulation enforces instance-level alignment, binding each image feature to its positive text anchor while repelling negatives.
It operationalizes the proposed semantic invariance principle: the representation must remain tethered to its true textual meaning even when visual distortions alter low-level appearance.
By discriminating at the pair level, the model preserves semantic identity under distortion without collapsing modalities into a single embedding space.

The discriminator is trained to classify $(F,E^+)$ and $(F',E^+)$ as real pairs and $(F,E^-)$ and $(F',E^-)$ as fake pairs:
\begin{equation}
\label{eq:disc_pair}
\begin{aligned}
\mathcal{L}_{\text{disc}}
&= \mathbb{E}_{(F,E^+),(F',E^+)}\!\Big[
\log D_{\psi}(F,E^+) + \log D_{\psi}(F',E^+)
\Big] \\
&\quad + \mathbb{E}_{(F,E^-),(F',E^-)}\!\Big[
\log\!\big(1 - D_{\psi}(F,E^-)\big) \\
&\quad + \log\!\big(1 - D_{\psi}(F',E^-)\big)
\Big].
\end{aligned}
\end{equation} 
During adversarial training, $f_{\theta}$ is optimized to fool $D_{\psi}$ by minimizing the opposite objective:
\begin{equation}
\label{eq:adv_pair}
\begin{aligned}
\mathcal{L}_{\text{adv}}
&= \mathbb{E}_{(F,E^+),(F',E^+)}\!\Big[
\log\!\big(1 - D_{\psi}(F,E^+)\big)\\
&\quad + \log\!\big(1 - D_{\psi}(F',E^+)\big)
\Big], 
\end{aligned}
\end{equation}
forcing image features to align tightly with the correct text anchors while remaining separable from negative ones.
The overall adversarial optimization alternates GAN-style updates:
\vspace{-0.85em}
\begin{align}
\max_{\psi}\;& \mathcal{L}_{\text{disc}}, 
\qquad
\min_{\theta}\; \mathcal{L}_{\text{feat}}
:= \lambda_{\text{adv}}\mathcal{L}_{\text{adv}}.
\label{eq:minmax_pair}
\end{align}
In each iteration, we (i) update $D_{\psi}$ using Eq.~\eqref{eq:disc_pair} to improve pair discrimination, and (ii) update $f_{\theta}$ using Eq.~\eqref{eq:adv_pair} to strengthen feature alignment with the positive text anchor. 
This pair-based adversarial training teaches the invariant extractor to preserve semantic meaning while ignoring distortion-specific cues, yielding image features that are both text-anchored and distortion-invariant.

\vspace{-1.5em}
\paragraph{Architecture.}
The feature extractor $f_{\theta}$ consists of a frozen CLIP image encoder followed by a trainable invariant feature extractor composed of three residual blocks and a projection head. 
Each residual block contains two linear layers with batch normalization, dropout, and a skip connection, formulated as 
$\mathbf{h}_{l+1} = \sigma\!\left(\text{BN}_2(W_2(\sigma(\text{BN}_1(W_1\mathbf{h}_l)))) + \mathbf{h}_l\right)$, 
where $\sigma(\cdot)$ denotes the ReLU activation. 
The projection head maps the output to a 1024-dimensional embedding, serving as the final invariant feature representation. 

The discriminator $D_{\psi}$ is implemented as a lightweight Transformer encoder that receives an image–text feature pair $(\mathbf{z}_I, \mathbf{z}_T)$ and predicts whether they correspond to the same semantic content. 
It comprises $L=4$ Transformer blocks with $H=8$ attention heads and a hidden dimension of 512, each including multi-head self-attention, feed-forward layers with GELU activation, and residual normalization. 
A learnable \texttt{[CLS]} token is prepended to the sequence of projected embeddings, $\mathbf{x} = [\mathbf{t}_{\text{cls}};\, W\mathbf{z}_I;\, W\mathbf{z}_T]$, and the output from the final \texttt{[CLS]} token is passed through a linear head to produce binary logits: 
$D_{\psi}(\mathbf{z}_I,\mathbf{z}_T) = \text{softmax}(W_o\,\mathbf{x}_{\text{cls}})$. 
This design balances efficiency and expressiveness, enabling the discriminator to capture fine-grained semantic correspondence between image and text pairs.

\vspace{-1.5em}
\paragraph{Auto-Augmentor and Feature Extractor Adversarial Training.}
The Auto-Augmentor $\mathcal{T}_{\text{aug}}(\cdot;\Theta)$ and feature extractor $f_{\theta}$ are trained in an adversarial fashion to jointly model realistic camera distortions and learn robust and invariant representations. 
Given an image $x$ and its distorted counterpart $\hat{x}=\mathcal{T}_{\text{aug}}(x;\Theta)$, the extractor produces feature embeddings $F=f_{\theta}(x)$ and $F'=f_{\theta}(\hat{x})$. 
The augmenter seeks to generate distortions that maximally disrupt feature consistency, while the extractor learns to minimize this discrepancy and restore semantic alignment. 
This process is formulated as a min–max optimization:
\begin{equation}
\label{eq:minmax_autoaug}
\min_{\theta}\;\max_{\Theta}\;
\mathcal{L}_{\text{inv}}(F',F)
-\lambda_{\text{sem}}\mathcal{L}_{\text{sem}}(x,\hat{x}),
\end{equation}
where $\mathcal{L}_{\text{inv}}$ measures the cosine dissimilarity between invariant features, and $\mathcal{L}_{\text{sem}}$ enforces perceptual fidelity through a frozen ViT encoder $E(\cdot)$:
\(
\mathcal{L}_{\text{inv}}(F',F)=1-\cos(F',F),
\)
and
\(
\mathcal{L}_{\text{sem}}(x,\hat{x})=\sum_i\!\big(1-\cos(E_i(x),E_i(\hat{x}))\big).
\)
During training, we alternate gradient \emph{ascent} on $\Theta$ to maximize 
$\mathcal{L}_{\text{inv}}-\lambda_{\text{sem}}\mathcal{L}_{\text{sem}}$, 
forcing the augmenter to discover increasingly challenging yet semantically faithful distortions, 
and gradient \emph{descent} on $\theta$ to minimize $\mathcal{L}_{\text{inv}}$, compelling $f_{\theta}$ to become invariant to those perturbations. 
To prevent trivial co-adaptation, gradients through $f_{\theta}$ are stopped during the augmenter step, and parameters in $\mathcal{T}_{\text{aug}}$ (e.g., blur radius, compression strength, gamma value) are clamped to physically plausible ranges.

As the Auto-Augmentor learns to generate progressively stronger camera-like distortions, the feature extractor $f_{\theta}$ simultaneously learns to neutralize them. 
This adversarial dynamic yields semantically stable and distortion-robust features that remain consistent even under complex capture conditions such as screen photographing, mobile recapture, or multi-stage compression.

\noindent\textbf{Full System Training.}
In the full framework, this Auto-Augmentor–Extractor adversarial loop operates jointly with the Cross-Modal Alignment between Image and Text representations. 
The overall system alternates among three coordinated updates: 
(i) training the discriminator $D_{\psi}$ to distinguish matched and mismatched image–text pairs, 
(ii) updating $\mathcal{T}_{\text{aug}}$ to produce adversarial yet semantically valid perturbations, and 
(iii) updating $f_{\theta}$ to align invariant image features with their correct text anchors while resisting camera distortions. 
This three-way optimization ensures that the learned representation is simultaneously semantically invariant, distortion-robust, and camera-realistic.

\vspace{-0.5em}
\subsection{Zero-Watermarking with Invariant Features}
\label{subsec:zero_wm}
\vspace{-0.5em}

The right block of Fig.~\ref{fig:proposed_method} illustrates a learning-based zero-watermarking head on top of our proposed invariant feature extractor $f_{\theta}$. We adopt the work on learning to associate zero-watermarks in feature space~\cite{tanvir2025invzw}, where the original image is never modified. Instead, we register a compact reference signature that binds a binary watermark message to the image’s invariant representation. At test time, the same frozen $f_{\theta}$ is applied to a (possibly camera-captured) image to extract the message.

\vspace{-1.5em}
\paragraph{Feature Encoding.}
Given an image $x$, we extract an invariant feature map
\(
F = f_{\theta}(x)\in\mathbb{R}^{H\times W\times C},
\)
and transform it to a vector via a lightweight head $\Psi$ (global average pooling followed by a linear projection):
\(
\tilde{F}=\Psi(F)\in\mathbb{R}^{d}.
\)
During watermark registration, $\theta$ is frozen; only the small head $\Psi$ and a reference codebook are optimized.

\vspace{-1.5em}
\paragraph{Reference Signature and Prediction.}
Let $W\in\{0,1\}^{k}$ denote the target $k$-bit message. We maintain a learnable reference matrix $C\in\mathbb{R}^{k\times d}$, whose $i$-th row $C_i$ acts as a directional code for bit $W_i$. The predicted bit probability is
\(
\hat{W}_i=\sigma\!\left(\tilde{F}\cdot C_i\right), i=1,\dots,k,
\)
where $\sigma(\cdot)$ is the sigmoid function.

\vspace{-1.5em}
\paragraph{Registration Objective.}
The registration is a small convex-like optimization over $C$ and the linear projection $\Psi$; inference reduces to a dot product and threshold. 
We optimize $C$ (and $\Psi$) by minimizing a binary cross-entropy with an $\ell_2$ regularizer on $C$:
\begin{align}
\label{eq:zw_loss}
\mathcal{L}_W &= -\sum_{i=1}^{k}\Big[W_i\log\hat{W}_i+(1-W_i)\log(1-\hat{W}_i)\Big],\\
\label{eq:zw_reg}
\mathcal{L}_C &= \|C\|_2^2,\qquad
\mathcal{L}_{\text{total}}=\mathcal{L}_W+\lambda_C \mathcal{L}_C,
\end{align}
with gradient updates
\begin{equation}
\label{eq:zw_update}
C \leftarrow C-\eta\,\frac{\partial \mathcal{L}_{\text{total}}}{\partial C},
\qquad
\Psi \leftarrow \Psi-\eta\,\frac{\partial \mathcal{L}_{\text{total}}}{\partial \Psi}.
\end{equation}
This optimization is performed per image–message pair: given a single image and a single $k$-bit watermark, we learn a dedicated reference signature $(C,\Psi)$ for that pair only: no minibatching, no modification to the image, and no updates to $f_{\theta}$.

\vspace{-1.5em}
\paragraph{Extraction.}
Given a potentially distorted (e.g., camera-captured) image $x'$, we compute
\begin{equation}
\label{eq:zw_extract}
\tilde{F}'=\Psi\!\big(f_{\theta}(x')\big),\qquad
\hat{W}'_i=\sigma\!\left(\tilde{F}'\cdot C_i\right),
\end{equation}
and recover the binary message via thresholding
\begin{equation}
\label{eq:zw_thresh}
\tilde{W}_i=\begin{cases}
1,& \hat{W}'_i\ge 0.5,\\
0,& \hat{W}'_i< 0.5~.
\end{cases}
\end{equation}
Robust recovery follows from the distortion invariance of $f_{\theta}$, which ensures $\tilde{F}'\approx \tilde{F}$ even after severe photometric, geometric, or compression artifacts.







\vspace{-0.5em}
\section{Experiment and Analysis}
\label{sec:experiment}
\vspace{-0.5em}

This section presents the empirical validation of the proposed TIACam framework. We organize the analysis into three parts. First, we evaluate the invariance and semantic quality of the learned features (Section~\ref{subsec:invariance_semantics}), demonstrating that our text-anchored representation remains stable under diverse distortions. Second, we assess the robustness of our watermarking system across both synthetic and camera-captured conditions (Section~\ref{subsec:camera_distortion}). Finally, we conduct ablation studies (Section~\ref{sec:ablation}) to analyze the roles of text guidance, and invariant feature learner.

\vspace{-0.5em}
\subsection{Experimental Setup}

\vspace{-0.5em}
\paragraph{Datasets.}
We conduct experiments on multiple datasets to evaluate both invariance learning and watermarking robustness. 
For text–image alignment, we jointly use Visual Genome~\cite{krishna2017visual} and Flickr30k~\cite{plummer2015flickr30k}, which together provide diverse textual granularity: region-level phrases from Visual Genome and full-sentence captions from Flickr30k. 
In addition, we test on ImageNet~\cite{deng2009imagenet}, MSCOCO~\cite{lin2014coco}, and Caltech-256~\cite{griffin2022caltech} to assess generalization across domains and resolutions. 
The proposed invariant feature extractor is trained on the combined Visual Genome and Flickr30k datasets. 
For Visual Genome, each region-level word set (e.g., \{person, wearing, fedora\}) is converted into a natural-language template,
\textit{``This is an image of a person wearing a fedora.''},
ensuring grammatical coherence and semantic alignment with the corresponding visual regions. 
This mixed training setup provides both localized and sentence-level supervision, enabling more stable multimodal feature learning.

\vspace{-1.5em}
\paragraph{Implementation Details.}
The framework follows the overall scheme illustrated in Fig.~\ref{fig:proposed_method}. 
It is implemented in PyTorch and trained on an NVIDIA RTX 4090 GPU. 
The learning rate is set to $1\times10^{-4}$ with the Adam optimizer. 
All images are resized to $128\times128\times3$.
Remarkably, the auto-augmentor modules are pretrained on their corresponding target distortions: geometric, photometric, additive noise, filtering, compression, and Moiré interference, using 10k random synthetic image pairs per type. 
Each learns its transformation via mean squared error (MSE) and structural similarity (SSIM) losses, and this pretraining provides meaningful initialization, ensuring physically plausible and semantically coherent updates. 
Subsequently, in adversarial training, the auto-augmentor is tuned to generate distortions that maximally perturb feature alignment, while the feature extractor $f_\theta$ learns to counteract them. 

\vspace{-0.5em}
\subsection{Feature Invariance and Semantic Evaluation}
\label{subsec:invariance_semantics}
\vspace{-0.5em}

We next evaluate the invariance and semantic quality of the learned representations. The analysis proceeds in two stages: (1) feature robustness, which quantifies the stability of representations under diverse distortions, and (2) semantic transferability, which measures how well the learned invariant features preserve discriminative meaning when used for downstream tasks.

\begin{table}[ht]
\centering
\vspace{-0.5em}
\scriptsize
\caption{Cosine similarity between features of original and distorted images.}
\vspace{-1.5em}
\resizebox{\columnwidth}{!}{%
\begin{tabular}{lcccccc}
\toprule
Distortion & SimCLR & BYOL & Barlow & VICReg & VIbCReg & TIACam \\
\midrule
Additive    & 0.82 & 0.88 & 0.79 & 0.83 & 0.89 & \textbf{0.97} \\
Photometric & 0.84 & 0.84 & 0.81 & 0.76 & 0.88 & \textbf{0.93} \\
Perspective & 0.87 & 0.85 & 0.87 & 0.83 & 0.88 & \textbf{0.95} \\
JPEG        & 0.79 & 0.80 & 0.87 & 0.81 & 0.73 & \textbf{0.98} \\
Moir\'e     & 0.85 & 0.83 & 0.84 & 0.89 & 0.87 & \textbf{0.97} \\
Filtering   & 0.88 & 0.88 & 0.89 & 0.87 & 0.88 & \textbf{0.98} \\
All         & 0.74 & 0.71 & 0.74 & 0.77 & 0.77 & \textbf{0.94} \\
\bottomrule
\end{tabular}%
}
\label{tab:cosine_ssl}
\vspace{-1.5em}
\end{table}

\vspace{-1.5em}
\paragraph{Feature Robustness.}
Table~\ref{tab:cosine_ssl} compares the cosine similarity between features of original and distorted images across six distortion types, averaged over the testing set of the evaluated datasets. 
TIACam consistently achieves the highest similarity values among self-supervised baselines, including SimCLR~\cite{chen2020simple}, BYOL~\cite{grill2020bootstrap}, Barlow Twins~\cite{zbontar2021barlow}, VICReg~\cite{bardes2021vicreg}, and VIbCReg~\cite{lee2021vibcreg}, indicating superior invariance to additive noise, photometric variation, perspective changes, and JPEG compression. 
Even under the compound distortion setting (\emph{All}), TIACam maintains stable representations with minimal degradation. 
These results verify that our text-anchored and auto-augmentation–guided training effectively enforces distortion-invariant feature learning.

\begin{figure}[ht]
\vspace{-0.5em}
\centering
\includegraphics[width=0.7\columnwidth]{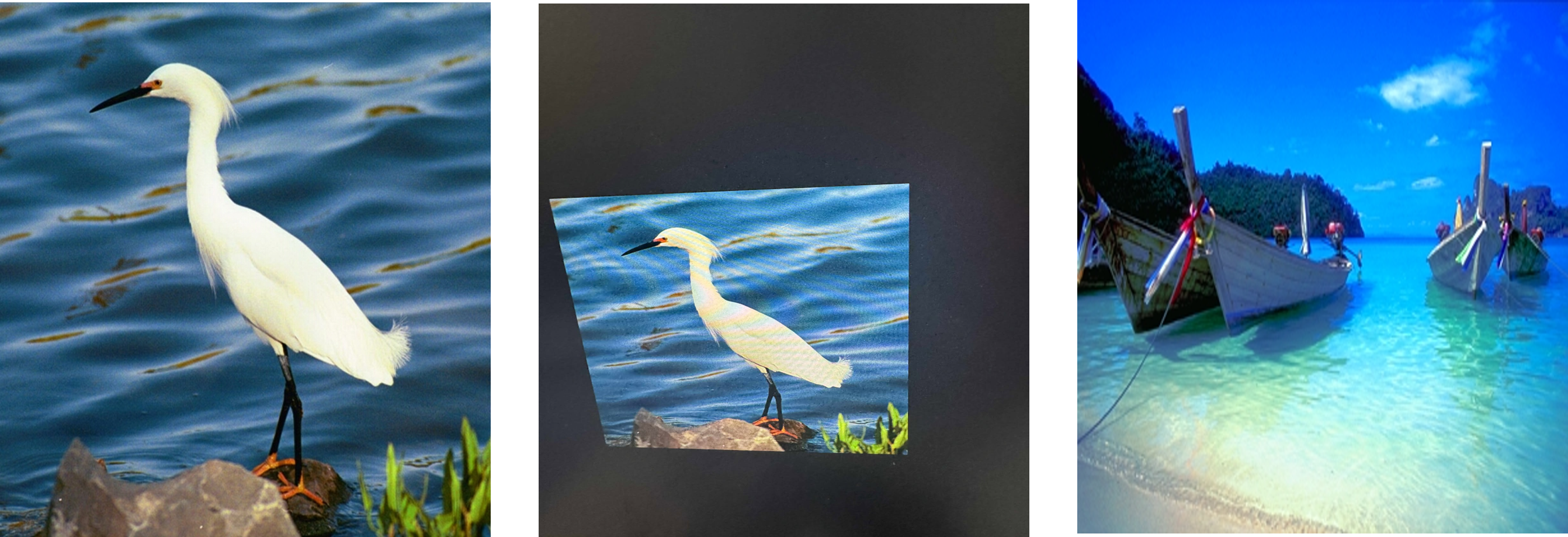}
\vspace{-0.5em}
\caption{From left to right: example image, its camera-distorted version, and an unrelated negative image.}
\label{fig:camera_distortion}
\vspace{-1.5em}
\end{figure}

Furthermore, the cosine similarity values can be interpreted intuitively: semantically consistent image pairs yield positive cosine values close to 1, whereas unrelated or mismatched pairs (even those visually similar in color or texture) produce negative values approaching –1. 
As shown in Fig.~\ref{fig:camera_distortion}, the proposed invariant feature space clearly distinguishes between true and false pairs, indicating strong semantic alignment. Across all testing data, positive image pairs consistently maintain high similarity (average $+98.4\%$), while negative pairs exhibit strong separation (average $-47.2\%$), confirming the discriminative stability of our learned features.

\begin{figure}[ht]
\centering
\includegraphics[width=1.0\columnwidth]{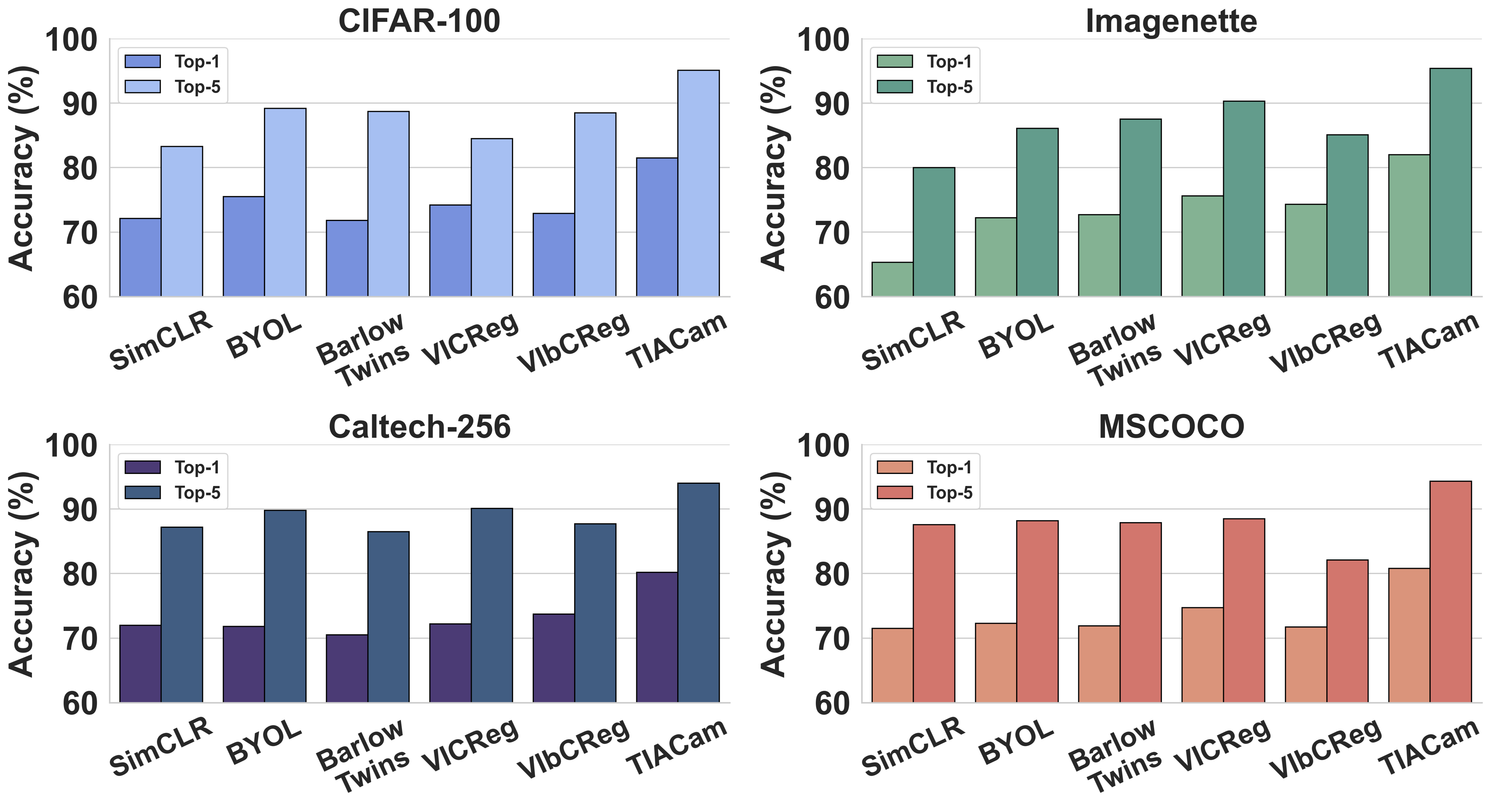}
\vspace{-1.5em}
\caption{Linear probe (Top 1 \& 5 accuracy) on CIFAR-100, Imagenette, MSCOCO, and Caltech-256 under six camera distortions.}
\label{fig:linear_eval}
\vspace{-2.0em}
\end{figure}

\vspace{-1.0em}
\paragraph{Semantic Transferability.}
To examine whether invariant features also preserve semantic meaning, we freeze the learned encoder and train a linear classifier probe on four benchmark datasets: CIFAR-100, Imagenette, MSCOCO, and Caltech-256 (15k train / 5k test). 
As reported in Figure~\ref{fig:linear_eval}, TIACam attains the highest Top-1 and Top-5 accuracies across all datasets and distortion types. 
For instance, on CIFAR-100 under additive and JPEG noise, TIACam improves Top-1 accuracy by over 7\% compared to the best baseline, and on large-scale datasets such as Caltech-256, it maintains over 80\% Top-1 accuracy even under joint distortions. 
This demonstrates that the proposed model not only learns invariant representations but also preserves their semantic discriminability, enabling robust generalization across content and domain variations.

\subsection{Watermarking under Camera Distortions}
\label{subsec:camera_distortion}
\vspace{-0.5em}

We next evaluate the robustness of \mn’s watermark extraction under real-world capture conditions, where optical artifacts and complex image transformations often degrade conventional watermarking systems. As illustrated in Fig.~\ref{fig:camera_distortion}, we consider three representative scenarios: screen camera capture, print camera capture, and screenshot (all with varied backgrounds), to comprehensively assess both camera-induced and user-generated distortions. 
Table~\ref{tab:camera_robustness} summarizes the results. 
Remarkbly, unlike prior methods~\cite{tancik2020stegastamp} that rely on a detection stage to localize the watermarked region before extraction, our approach directly processes the entire image. Owing to the high robustness of the learned invariant feature space, \mn\ achieves reliable extraction without explicit localization. This design choice simplifies deployment and still delivers state-of-the-art accuracy in challenging real cases; incorporating a detection step would further enhance performance.


\begin{figure*}[ht]
\centering
\includegraphics[width=0.95\linewidth]{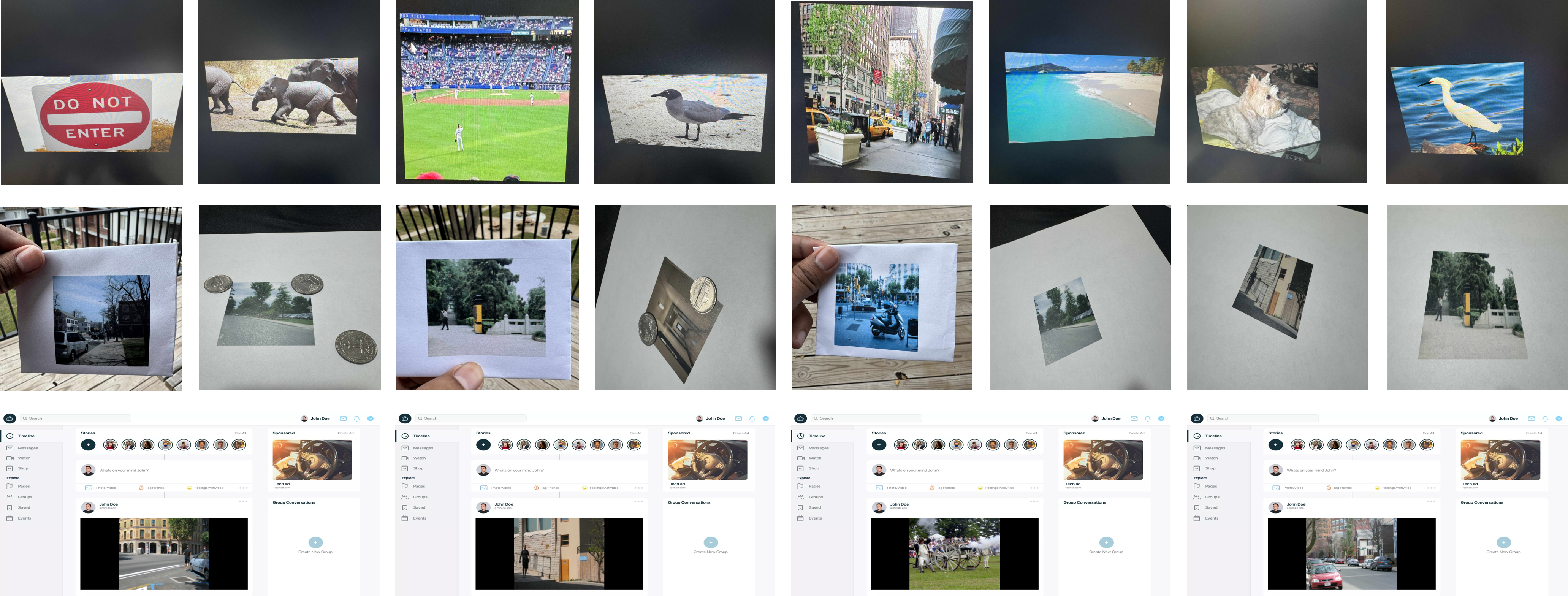}
\vspace{-0.5em}
\caption{Examples of real-world distortions evaluated in this study. 
The first row shows screen-camera captures, 
where images are re-photographed from different monitors using different mobile cameras. 
The second row shows print-camera captures, 
obtained by printing images on paper and re-capturing them under varying lighting and viewpoints. 
The third row shows screenshots,  
representing user-generated distortions such as compositional clutter. 
}

\label{fig:camera}
\vspace{-1.0em}
\end{figure*}

\vspace{-1.5em}
\paragraph{Screen Camera.}
We first evaluate the prevalent case of display recapture, where watermarked images are displayed on a monitor and re-captured using a mobile camera. This setting introduces complex degradations including perspective warping, illumination shifts, sensor noise, and color imbalance. Evaluated on over three hundreds of screen-captured testing images, TIACam achieves nearly perfect extraction with 99\% and 98\% bit accuracy for 30- and 100-bit messages, respectively. Competing methods such as HiDDeN~\cite{zhu2018hidden}, PIMoG~\cite{fang2022pimog}, and StegaStamp~\cite{tancik2020stegastamp} show substantial degradation, highlighting that TIACam’s feature-space invariance provides robustness beyond pixel-level watermarking.

\vspace{-1.5em}
\paragraph{Print Camera.}
We next examine robustness under real physical recapture, where printed images are re-photographed under diverse lighting and viewpoint conditions. Across over 200 printed-and-recaptured samples, TIACam maintains accurate recovery: 96.6\% and 95.1\% for 30- and 100-bit messages, respectively, significantly outperforming all baselines. These results demonstrate that the proposed \mn features preserve alignment across imaging pipelines, ensuring consistent watermark integrity even after multiple optical transformations.

\vspace{-1.5em}
\paragraph{Screenshots.}
To further assess resilience to user-generated edits, we applied the augmentation strategy of~\cite{fernandez2022watermarking} to a large set of screenshots and meme-style crops with varied backgrounds. TIACam achieves 97.4\% and 95.2\% bit accuracy for 30- and 100-bit messages, respectively, demonstrating exceptional robustness against uncontrolled distortions and semantic edits. Even without region detection, the model successfully extracts watermarks from entire images, confirming the discriminative power of our invariant feature space.

\begin{table}[ht]
\vspace{-0.5em}
\centering
\caption{Bit accuracy (BA, in \%) comparison of HiDDeN, PIMoG, StegaStamp, and TIACam under three real-world distortion types: Screen Camera, Print Camera, and Screenshots, for 30-bit and 100-bit message lengths. Results averaged over three random seeds; variance below 0.3\%.}
\vspace{-1.0em}
\resizebox{\columnwidth}{!}{
\begin{tabular}{lcccccc}
\toprule
\multirow{2}{*}{\textbf{Method}} 
& \multicolumn{2}{c}{\textbf{Screen Camera}} 
& \multicolumn{2}{c}{\textbf{Print Camera}} 
& \multicolumn{2}{c}{\textbf{Screenshots}} \\
\cmidrule(lr){2-3} \cmidrule(lr){4-5} \cmidrule(lr){6-7}
& 30 bits & 100 bits & 30 bits & 100 bits & 30 bits & 100 bits \\
\midrule
HiDDeN      & 70.6\% & 68.8\% & 67.1\% & 65.7\% & 74.5\% & 70.6\% \\
PIMoG       & 82.3\% & 80.1\% & 75.7\% & 72.3\% & 79.7\% & 78.6\% \\
StegaStamp  & 93.8\% & 91.2\% & 92.2\% & 91.3\% & 93.7\% & 93.9\% \\
\textbf{TIACam} & \textbf{99.1\%} & \textbf{98.2\%} & \textbf{96.6\%} & \textbf{95.1\%} & \textbf{97.4\%} & \textbf{95.2\%} \\
\bottomrule
\end{tabular}
}
\label{tab:camera_robustness}
\vspace{-1.5em}
\end{table}

\vspace{-0.75em}
\subsection{Ablation Studies}
\label{sec:ablation}
\vspace{-0.5em}

\paragraph{Effectiveness of \mn Feature Extractor.}
We perform ablation experiments to validate that the robustness of \mn does not originate from the pretrained CLIP backbone but from our proposed invariant feature learning framework. Specifically, we compare (1) CLIP Baseline, which directly uses the CLIP image encoder, and (2) CLIP + \mn Invariant Feature Extractor, which incorporates our proposed refinement module trained under adversarial auto-augmentation. 
We evaluate feature robustness on 10k testing images from each of four datasets: Visual Genome, Flickr, MSCOCO, and ImageNet, under all six distortions randomly sampled from the camera-distortion pipeline. The average cosine similarity between features of original and distorted images is reported in Table~\ref{tab:ablation_clip_feature}. \mn substantially enhances stability, improving cosine similarity by approximately 13–15\% across datasets. This confirms that the learned invariance arises from our framework rather than from CLIP’s pretrained representations.

\begin{table}[ht]
\vspace{-0.5em}
\centering
\caption{Ablation on \mn feature extractor. Higher cosine similarity indicates stronger invariance under distortion.}
\vspace{-1.0em}
\resizebox{\columnwidth}{!}{
\begin{tabular}{lcc}
\hline
\textbf{Dataset} & \textbf{CLIP Only} & \textbf{CLIP + \mn Feature} \\
\hline
Visual Genome & 0.78 & \textbf{0.92} \\
Flickr        & 0.84 & \textbf{0.93} \\
MSCOCO        & 0.76 & \textbf{0.89} \\
ImageNet      & 0.82 & \textbf{0.93} \\
\hline
\end{tabular}
}
\label{tab:ablation_clip_feature}
\vspace{-1.5em}
\end{table}

\begin{figure}[ht]
\vspace{-1.0em}
\centering
\includegraphics[width=0.5\columnwidth]{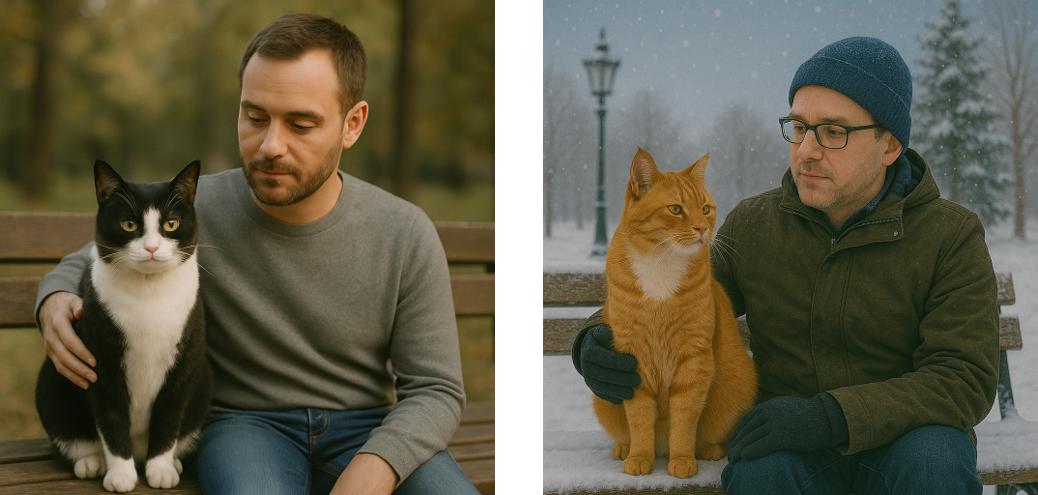}
\vspace{-0.5em}
\caption{Extreme case test of feature distinctiveness. 
Both images are generated with the same caption, \textit{``A photo of cat, person, and park bench''}, yet their invariant features exhibit a cosine similarity of only 0.73.}
\label{fig:same_caption_case}
\vspace{-1.0em}
\end{figure}

\vspace{-1.0em}
\paragraph{Distinctiveness of \mn\ Image Feature Space.}
To examine the balance between invariance and distinctiveness, we design extreme cases where two visually different images share the same caption, such as \textit{``A photo of cat, person, and park bench''} in Fig.~\ref{fig:same_caption_case}. 
We embed a 100-bit watermark using Image~1 and attempt extraction from Image~2 and from the text embedding directly. 

To quantify this phenomenon, we generate 200 image pairs using text-to-image synthesis with identical captions. 
Only Image~1 yields perfect recovery (100\% bit accuracy), while extraction from Image~2 and text features drops to an average of 84\%. 
The average cosine similarity between Image 1 and 2 invariant features is 0.73, indicating that our model maintains strong feature distinctiveness even under identical text descriptions. 
During training, the text anchors have mixed granularity, which encourages the model to learn both invariance to distortions and separability across distinct visual instances.
These results confirm that \mn\ effectively balances semantic invariance and visual individuality: achieving text-guided consistency without collapsing visually distinct samples.

\vspace{-1.0em}
\section{Conclusion}
\label{sec:conclusion}
\vspace{-0.5em}

We presented \mn, a framework for camera-robust zero-watermarking. 
It features a novel learnable auto-augmentor that models camera-like distortions, 
a novel text-anchored invariant feature extractor trained in a three-way adversarial loop, 
and demonstrated state-of-the-art robustness across real camera and related distortions.

\newpage
\vspace{-1em}
{
    \small
    \bibliographystyle{ieeenat_fullname}
    \bibliography{reference}

@String(CVPR= {IEEE Conf. Comput. Vis. Pattern Recog.})

@String(ICCV= {Int. Conf. Comput. Vis.})

@String(ECCV= {Eur. Conf. Comput. Vis.})

@String(ICASSP=	{ICASSP})

@String(ICLR = {Int. Conf. Learn. Represent.})

@String(AAAI = {AAAI})

@String(CVPR  = {CVPR})

@String(ICCV  = {ICCV})

@String(ECCV  = {ECCV})

@String(ICLR  = {ICLR})

@article{li2024zwnet,
  title        = {{ZWNet: A Deep-Learning-Powered Zero-Watermarking Scheme With High Robustness and Discriminability for Images}},
  author       = {Li, Can and Sun, Hua and Wang, Changhong and Chen, Sheng and Liu, Xi and Zhang, Yi and Ren, Na and Tong, Deyu},
  journal      = {Applied Sciences},
  volume       = {14},
  number       = {1},
  pages        = {435},
  year         = {2024}
}

@inproceedings{fierro2019robust,
  title        = {{A Robust Image Zero-Watermarking Using Convolutional Neural Networks}},
  author       = {Fierro-Radilla, Atoany and Nakano-Miyatake, Mariko and Cedillo-Hernandez, Manuel and Cleofas-Sanchez, Laura and Perez-Meana, Hector},
  booktitle    = {IWBF},
  pages        = {1--5},
  year         = {2019}
}

@inproceedings{fang2022pimog,
  title        = {{PIMoG: An Effective Screen-Shooting Noise-Layer Simulation for Deep-Learning-Based Watermarking Network}},
  author       = {Fang, Han and Jia, Zhaoyang and Ma, Zehua and Chang, Ee-Chien and Zhang, Weiming},
  booktitle    = {ACM MM},
  pages        = {2267--2275},
  year         = {2022}
}

@inproceedings{guo2023practical,
  title        = {{Practical Deep Dispersed Watermarking With Synchronization and Fusion}},
  author       = {Guo, Hengchang and Zhang, Qilong and Luo, Junwei and Guo, Feng and Zhang, Wenbin and Su, Xiaodong and Li, Minglei},
  booktitle    = {Proceedings of the 31st ACM International Conference on Multimedia},
  pages        = {7922--7932},
  year         = {2023}
}

@inproceedings{jia2021mbrs,
  title        = {{MBRS: Enhancing Robustness of DNN-Based Watermarking by Mini-Batch of Real and Simulated JPEG Compression}},
  author       = {Jia, Zhaoyang and Fang, Han and Zhang, Weiming},
  booktitle    = {ACM MM},
  pages        = {41--49},
  year         = {2021}
}

@inproceedings{tancik2020stegastamp,
  title        = {{StegaStamp: Invisible Hyperlinks in Physical Photographs}},
  author       = {Tancik, Matthew and Mildenhall, Ben and Ng, Ren},
  booktitle    = {CVPR},
  pages        = {2117--2126},
  year         = {2020}
}

@inproceedings{wang2024must,
  title        = {{MUST: Robust Image Watermarking for Multi-Source Tracing}},
  author       = {Wang, Guanjie and Ma, Zehua and Liu, Chang and Yang, Xi and Fang, Han and Zhang, Weiming and Yu, Nenghai},
  booktitle    = {AAAI},
  pages        = {5364--5371},
  year         = {2024}
}

@inproceedings{zhu2018hidden,
  title        = {{HiDDeN: Hiding Data With Deep Networks}},
  author       = {Zhu, Jiren and Kaplan, Russell and Johnson, Justin and Fei-Fei, Li},
  booktitle    = {ECCV},
  pages        = {657--672},
  year         = {2018}
}

@article{vukotic2020classification,
  title        = {{Are Classification Deep Neural Networks Good for Blind Image Watermarking?}},
  author       = {Vukoti{\'c}, Vedran and Chappelier, Vivien and Furon, Teddy},
  journal      = {Entropy},
  volume       = {22},
  number       = {2},
  pages        = {198},
  year         = {2020}
}

@inproceedings{fernandez2022watermarking,
  title        = {{Watermarking Images in Self-Supervised Latent Spaces}},
  author       = {Fernandez, Pierre and Sablayrolles, Alexandre and Furon, Teddy and J{\'e}gou, Herv{\'e} and Douze, Matthijs},
  booktitle    = {ICASSP 2022 IEEE International Conference on Acoustics, Speech and Signal Processing (ICASSP)},
  pages        = {3054--3058},
  year         = {2022}
}

@inproceedings{khatri2022perspective,
  title        = {{Perspective Transformation Layer}},
  author       = {Khatri, Nishan and Dasgupta, Agnibh and Shen, Yucong and Zhong, Xin and Shih, Frank Y.},
  booktitle    = {2022 International Conference on Computational Science and Computational Intelligence (CSCI)},
  pages        = {1395--1401},
  year         = {2022}
}

@inproceedings{jang2017categorical,
  title        = {{Categorical Reparametrization With Gumbel-Softmax}},
  author       = {Jang, Eric and Gu, Shixiang and Poole, Ben},
  booktitle    = {International Conference on Learning Representations (ICLR 2017)},
  year         = {2017}
}

@misc{semenov2025smooth,
  author       = {Semenov, Stanislav},
  title        = {{Smooth Approximations of the Rounding Function}},
  howpublished = {\url{https://arxiv.org/abs/2504.19026}},
  year         = {2025},
  note         = {arXiv Preprint}
}

@inproceedings{tishby2015deep,
  title        = {{Deep Learning and the Information Bottleneck Principle}},
  author       = {Tishby, Naftali and Zaslavsky, Noga},
  booktitle    = {2015 IEEE Information Theory Workshop (ITW)},
  pages        = {1--5},
  year         = {2015}
}

@misc{tanvir2025invzw,
  author       = {Tanvir, Abdullah All and Zhong, Xin},
  title        = {{InvZW: Invariant Feature Learning via Noise-Adversarial Training for Robust Image Zero-Watermarking}},
  howpublished = {\url{https://arxiv.org/abs/2506.20370}},
  year         = {2025},
  note         = {arXiv Preprint}
}

@misc{ahtesham2025text,
  author       = {Ahtesham, Muhammad and Zhong, Xin},
  title        = {{Text-Guided Image Invariant Feature Learning for Robust Image Watermarking}},
  howpublished = {\url{https://arxiv.org/abs/2503.13805}},
  year         = {2025},
  note         = {arXiv Preprint}
}

@inproceedings{liu2025watermarking,
  title        = {{Watermarking One for All: A Robust Watermarking Scheme Against Partial Image Theft}},
  author       = {Liu, Gaozhi and Cao, Silu and Qian, Zhenxing and Zhang, Xinpeng and Li, Sheng and Peng, Wanli},
  booktitle    = {Proceedings of the IEEE/CVF Conference on Computer Vision and Pattern Recognition},
  pages        = {8225--8234},
  year         = {2025}
}

@inproceedings{caron2021emerging,
  title        = {{Emerging Properties in Self-Supervised Vision Transformers}},
  author       = {Caron, Mathilde and Touvron, Hugo and Misra, Ishan and J{\'e}gou, Herv{\'e} and Mairal, Julien and Bojanowski, Piotr and Joulin, Armand},
  booktitle    = {ICCV},
  pages        = {9650--9660},
  year         = {2021}
}

@article{han2024application,
  title={Application of Zero-Watermarking Scheme Based on Swin Transformer for Securing the Metaverse Healthcare Data},
  author={Han, Baoru and Wang, Han and Qiao, Dawei and Xu, Jia and Yan, Tianyu},
  journal={IEEE journal of biomedical and health informatics},
  volume={28},
  number={11},
  pages={6360--6369},
  year={2024}
}

@article{he2023shrinkage,
  title        = {{Shrinkage and Redundant Feature Elimination Network-Based Robust Image Zero-Watermarking}},
  author       = {He, Lingqiang and He, Zhouyan and Luo, Ting and Song, Yang},
  journal      = {Symmetry},
  volume       = {15},
  number       = {5},
  pages        = {964},
  year         = {2023}
}

@inproceedings{deng2009imagenet,
  title        = {{ImageNet: A Large-Scale Hierarchical Image Database}},
  author       = {Deng, Jia and Dong, Wei and Socher, Richard and Li, Li-Jia and Li, Kai and Fei-Fei, Li},
  booktitle    = {2009 IEEE Conference on Computer Vision and Pattern Recognition},
  pages        = {248--255},
  year         = {2009}
}

@misc{lin2014coco,
  author       = {Lin, Tsung-Yi and Maire, Michael and Belongie, Serge J. and Bourdev, Lubomir D. and Girshick, Ross B. and Hays, James and Perona, Pietro and Ramanan, Deva and Doll{\'a}r, Piotr and Zitnick, C. Lawrence},
  title        = {{Microsoft COCO: Common Objects in Context}},
  howpublished = {\url{https://arxiv.org/abs/1405.0312}},
  year         = {2014},
  note         = {arXiv Preprint}
}

@article{krishna2017visual,
  title        = {{Visual Genome: Connecting Language and Vision Using Crowdsourced Dense Image Annotations}},
  author       = {Krishna, Ranjay and Zhu, Yuke and Groth, Oliver and Johnson, Justin and Hata, Kenji and Kravitz, Joshua and Chen, Stephanie and Kalantidis, Yannis and Li, Li-Jia and Shamma, David A.},
  journal      = {International Journal of Computer Vision},
  volume       = {123},
  number       = {1},
  pages        = {32--73},
  year         = {2017}
}

@inproceedings{plummer2015flickr30k,
  title={Flickr30k entities: Collecting region-to-phrase correspondences for richer image-to-sentence models},
  author={Plummer, Bryan A and Wang, Liwei and Cervantes, Chris M and Caicedo, Juan C and Hockenmaier, Julia and Lazebnik, Svetlana},
  booktitle={Proceedings of the IEEE international conference on computer vision},
  pages={2641--2649},
  year={2015}
}

@article{griffin2022caltech,
  title={Caltech 256 (1.0)[Data set]},
  author={Griffin, G and Holub, A and Perona, P},
  journal={CaltechDATA. doi},
  volume={10},
  pages={D1},
  year={2022}
}

@inproceedings{chen2020simple,
  title        = {{A Simple Framework for Contrastive Learning of Visual Representations}},
  author       = {Chen, Ting and Kornblith, Simon and Norouzi, Mohammad and Hinton, Geoffrey},
  booktitle    = {ICML},
  pages        = {1597--1607},
  year         = {2020}
}

@article{grill2020bootstrap,
  title        = {{Bootstrap Your Own Latent: A New Approach to Self-Supervised Learning}},
  author       = {Grill, Jean-Bastien and Strub, Florian and Altch{\'e}, Florent and Tallec, Corentin and Richemond, Pierre and Buchatskaya, Elena and Doersch, Carl and Avila Pires, Bernardo and Guo, Zhaohan and Gheshlaghi Azar, Mohammad},
  journal      = {Advances in Neural Information Processing Systems},
  volume       = {33},
  pages        = {21271--21284},
  year         = {2020}
}

@inproceedings{zbontar2021barlow,
  title        = {{Barlow Twins: Self-Supervised Learning via Redundancy Reduction}},
  author       = {Zbontar, Jure and Jing, Li and Misra, Ishan and LeCun, Yann and Deny, St{\'e}phane},
  booktitle    = {International Conference on Machine Learning},
  pages        = {12310--12320},
  year         = {2021}
}

@misc{bardes2021vicreg,
  author       = {Bardes, Adrien and Ponce, Jean and LeCun, Yann},
  title        = {{ViCReg: Variance-Invariance-Covariance Regularization for Self-Supervised Learning}},
  howpublished = {\url{https://arxiv.org/abs/2105.04906}},
  year         = {2021},
  note         = {arXiv Preprint}
}

@misc{lee2021vibcreg,
  title        = {{VIBCReg: Variance-Invariance-Better-Covariance Regularization for Self-Supervised Learning on Time Series}},
  author       = {Lee, Daesoo and Aune, Erlend},
  howpublished = {\url{https://arxiv.org/abs/2109.00783}},
  year         = {2021},
  note         = {arXiv Preprint}
}
}

\clearpage
\setcounter{page}{1}
\setcounter{figure}{0}
\setcounter{table}{0}
\maketitlesupplementary

\subsection*{S1. Training Data Diversity and Caption Examples}

\begin{figure*}[ht]
\vspace{-0.5em}
\centering
\includegraphics[width=1.0\linewidth]{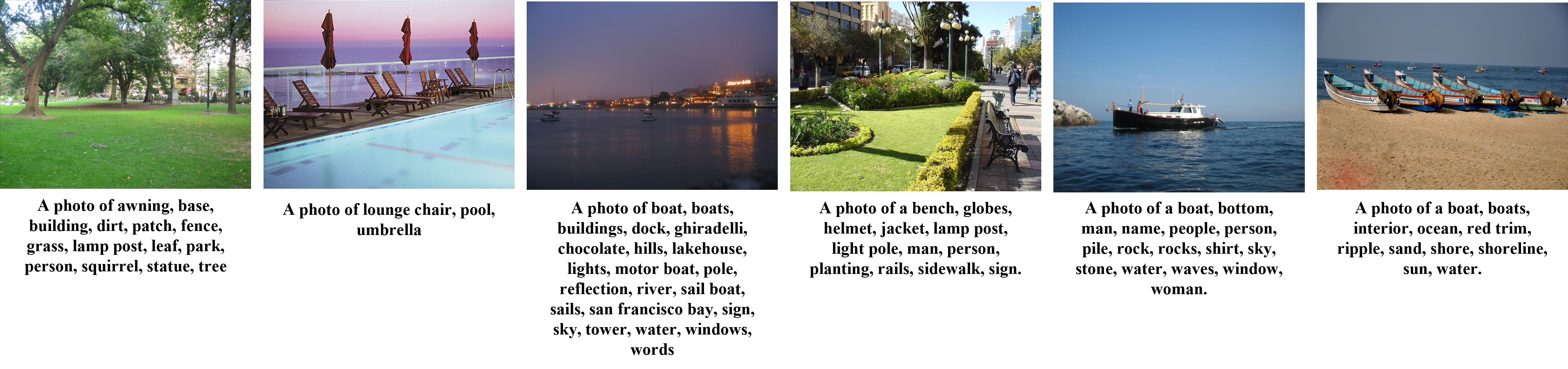}
\vspace{-1.75em}
\caption{Example of Visual Gnome Training Data used in TIACam.}
\label{fig:VG_eg}
\vspace{-0.25em}
\end{figure*}

\begin{figure*}[ht]
\vspace{-0.25em}
\centering
\includegraphics[width=1.0\linewidth]{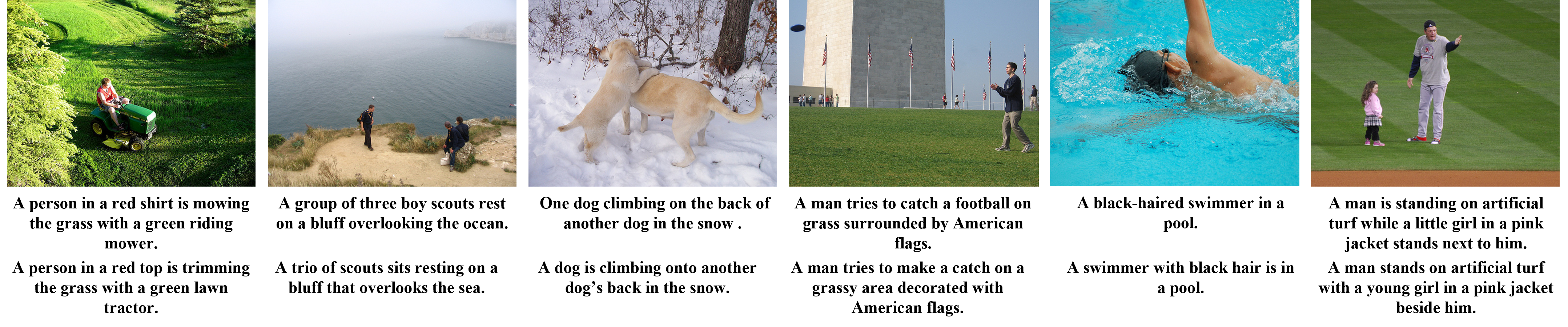}
\vspace{-1.5em}
\caption{Example of Flickr training data used in TIACam. The first row shows the images, the second row contains the corresponding captions, and the third row presents paraphrased versions of those captions.}
\label{fig:FL_eg}
\vspace{-1.0em}
\end{figure*}

To illustrate the diversity of text-image supervision used during training, we provide qualitative examples from the Visual Genome and Flickr30k datasets in Fig.~\ref{fig:VG_eg} and Fig.~\ref{fig:FL_eg}. Visual Genome contains short, object-centric region descriptions, often listing entities or attributes, while Flickr30k provides full natural sentences with richer structure and contextual detail. Despite this difference in caption granularity and linguistic style, TIACam trains stably on the combined dataset and produces consistent invariant features. This demonstrates that the model is not tied to a particular caption format and that its learned feature space generalizes across heterogeneous annotation styles, helping reduce dataset-specific bias and improving robustness in multimodal settings.

\section*{S2. Realism of the Learned Auto-Augmentor Distortions}

\begin{figure*}[ht]
\centering
\includegraphics[width=0.95\linewidth]{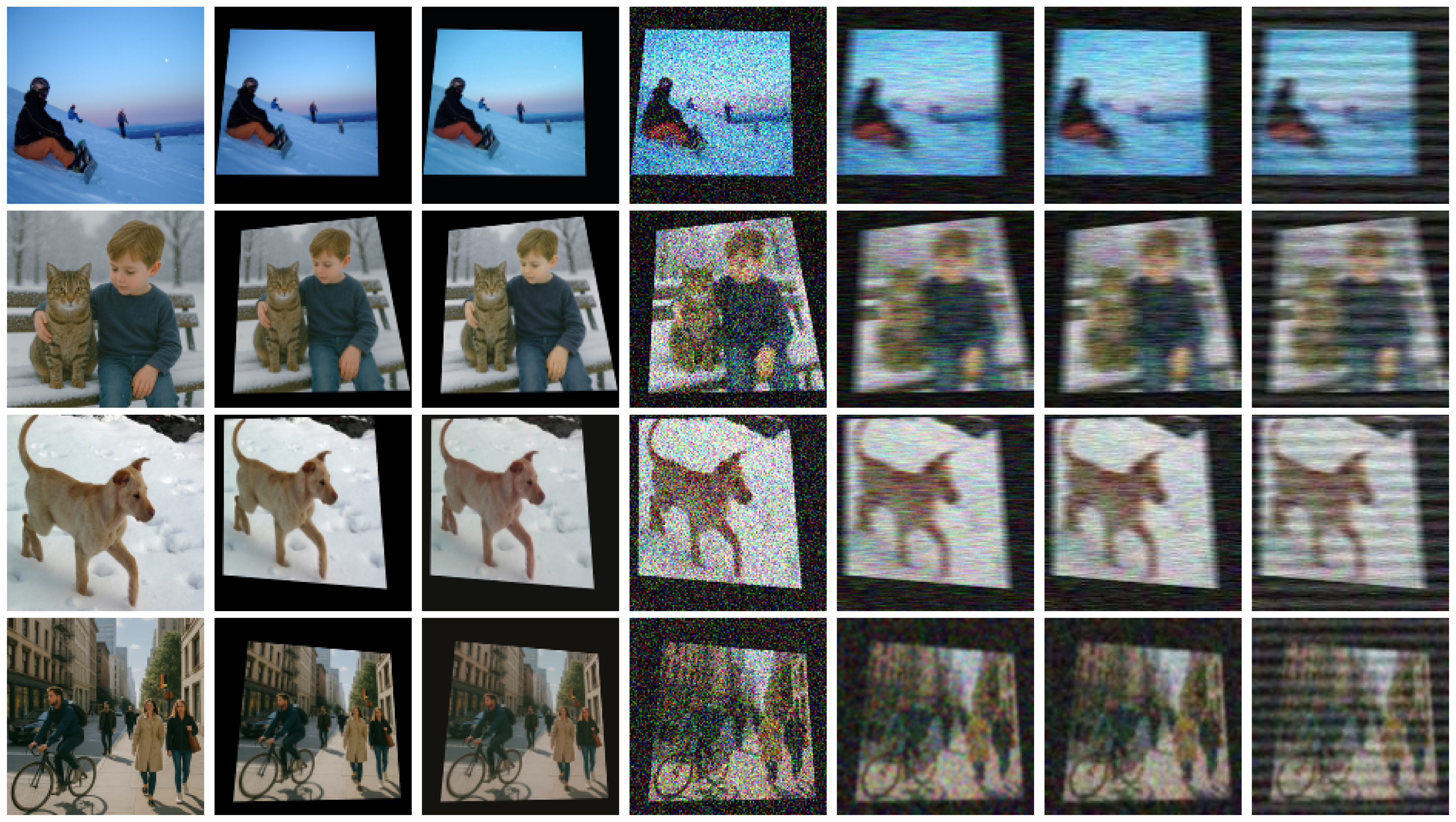}
\vspace{-0.7em}
\caption{Examples of auto-augmented images. }
\label{fig:more_augmentation}
\vspace{-1.0em}
\end{figure*}

To illustrate the realism and diversity of the distortions learned by our auto-augmentor, Fig.~\ref{fig:more_augmentation} presents representative examples. Each row shows the original input followed by distortions synthesized by the learned modules, including perspective deformation, photometric shifts, additive noise, filtering artifacts, JPEG degradation, and Moiré interference.

These visualizations show that the auto-augmentor produces distortions characteristic of real camera pipelines: such as chromatic imbalance, sensor noise, edge warping, and Moiré patterns, rather than simple digital corruptions. The close resemblance between these synthesized distortions and actual screen- and print-recapture effects supports our claim that the learned augmentations effectively model camera perturbations, enabling \mn\ to learn invariance that transfers robustly to real-world camera scenarios.

\section*{S3. Additional Ablation Study: Learned Auto-Augmentor vs.\ Manual Distortions}

To assess the impact of the learnable auto-augmentor, we replace it with a fixed set of hand-crafted distortions while keeping the feature extractor and training objectives unchanged. We then compare cosine similarity between invariant features from original and distorted images. As shown in Tables~\ref{tab:cosine_auto_vs_manual_6} and \ref{tab:cosine_auto_vs_manual_3}, the manual augmentor produces noticeably lower similarity across both synthetic distortions and real capture settings (screen camera, print camera, and screenshots). In contrast, the learned auto-augmentor consistently achieves much higher similarity, indicating stronger distortion-invariant feature alignment.

\begin{table}[ht]
\vspace{-0.5em}
\centering
\caption{Cosine similarity between original and distorted invariant features across six distortion types.}
\label{tab:cosine_auto_vs_manual_6}
\vspace{-0.75em}
\resizebox{\columnwidth}{!}{
\begin{tabular}{lcc}
\toprule
\textbf{Distortion Type} & \textbf{Manual Augmentor (Fixed)} & \textbf{Auto-Augmentor (Ours)} \\
\midrule
Additive Noise      & 0.84 & \textbf{0.98} \\
Photometric         & 0.86 & \textbf{0.96} \\
Perspective         & 0.78 & \textbf{0.92} \\
JPEG Compression    & 0.88 & \textbf{0.97} \\
Moir\'e Pattern     & 0.76 & \textbf{0.89} \\
Filtering & 0.89 & \textbf{0.98} \\
\bottomrule
\end{tabular}
}
\vspace{-1.0em}
\end{table}

\begin{table}[ht]
\vspace{-0.5em}
\centering
\caption{Cosine similarity between original and distorted invariant features across Screen Camera, Print Camera, Screenshot}
\label{tab:cosine_auto_vs_manual_3}
\vspace{-0.75em}
\resizebox{\columnwidth}{!}{
\begin{tabular}{lcc}
\toprule
\textbf{Distortion Type} & \textbf{Manual Augmentor (Fixed)} & \textbf{Auto-Augmentor (Ours)} \\
\midrule
Screen Camera      & 0.82 & \textbf{0.91} \\
Print Camera        & 0.76 & \textbf{0.87} \\
Screenshot        & 0.78 & \textbf{0.85} \\
\bottomrule
\end{tabular}
}
\vspace{-1.0em}
\end{table}

We further compare watermark extraction performance under real camera distortions. Table~\ref{tab:auto_vs_manual} reports bit accuracy for 30-, 100-, and 200-bit payloads under randomly selected six distortions out of Additive, Photometric, Perspective, JPEG, Moir\'e, and Filtering Noise. Consistent with the feature-level results, models trained with manual distortions degrade as the message length increases, while the learned auto-augmentor maintains high accuracy across all payload sizes. 

\begin{table}[ht]
\vspace{-1.0em}
\centering
\caption{Bit accuracy (\%) under randomly applied 6 distortions.}
\label{tab:auto_vs_manual}
\vspace{-0.75em}
\resizebox{0.75\columnwidth}{!}{
\begin{tabular}{lccc}
\toprule
\textbf{Model Variant} & \textbf{30-bit} & \textbf{100-bit} & \textbf{200-bit} \\
\midrule
Manual Augmentor (Fixed) & 91.2 & 89.5 & 89.3 \\
Auto-Augmentor (Ours)    & \textbf{98.7} & \textbf{96.4} & \textbf{96.1} \\
\bottomrule
\end{tabular}
}
\vspace{-0.5em}
\end{table}

Table \ref{tab:wm_manual_auto} provides a breakdown by evaluating the same two models under three specific capture pipelines: screen camera, print camera, and screenshots. The manual augmentor exhibits notable drops, especially under print-camera and high-bit scenarios, while the auto-augmentor consistently achieves strong accuracy across all conditions.

\begin{table}[ht]
\vspace{-0.5em}
\centering
\caption{Bit accuracy (\%) of Manual Augmentor and Auto Augmentor under screen camera, print camera, and screenshot distortions (30-bit and 100-bit).}
\vspace{-0.75em}
\resizebox{\columnwidth}{!}{
\begin{tabular}{lcccccc}
\toprule
\multirow{2}{*}{\textbf{Method}} 
& \multicolumn{2}{c}{\textbf{Screen Camera}} 
& \multicolumn{2}{c}{\textbf{Print Camera}} 
& \multicolumn{2}{c}{\textbf{Screenshots}} \\
\cmidrule(lr){2-3} \cmidrule(lr){4-5} \cmidrule(lr){6-7}
& 30 bits & 100 bits & 30 bits & 100 bits & 30 bits & 100 bits \\
\midrule
Manual Augmentor     & 93.3\% & 92.6\% & 90.1\% & 87.4\% & 92.7\% & 87.9\% \\
\textbf{Auto Augmentor (Ours)} 
& \textbf{99.1\%} & \textbf{98.4\%} 
& \textbf{96.9\%} & \textbf{95.3\%} 
& \textbf{97.6\%} & \textbf{96.5\%} \\
\bottomrule
\end{tabular}}
\label{tab:wm_manual_auto}
\vspace{-0.5em}
\end{table}

These results show that the learned auto-augmentor provides stronger and more realistic distortion modeling than fixed operators. By better approximating physical camera artifacts, the auto-augmentor enables \mn\ to learn more stable invariant features and substantially improves real-world watermark robustness.

\section*{S4. Comprehensive Comparison with Existing Digital and Zero-Watermarking Methods}

The main paper focuses on camera-based evaluations, where \mn\ shows clear advantages over learning-based watermarking baselines. For completeness, we further compare \mn\ with a broader set of digital watermarking and zero-watermarking systems under digital appearance distortions, geometric transformations, and real camera capture. These supplementary results confirm that the robustness of \mn\ extends well beyond the camera setting.

Table~\ref{tab:bit_accuracy} reports bit accuracy under a range of common digital perturbations. While pretrained feature-based zero-watermarking approaches remain reliable under mild settings, they degrade substantially under geometric or severe appearance changes such as rotation, heavy cropping, blur, and screenshot artifacts. In contrast, \mn\ maintains consistently high accuracy across all digital distortions, highlighting the benefit of learning distortion-invariant features rather than relying solely on pretrained representations.

\begin{table}[ht]
\centering
\caption{Bit accuracy (\%) under digital distortions.}
\vspace{-0.75em}
\label{tab:bit_accuracy}
\resizebox{0.65\linewidth}{!}{%
\begin{tabular}{|l|c|c|c|}
\hline
\textbf{Distortion} & \textbf{ZBW \cite{vukotic2020classification}} & \textbf{WSSL \cite{fernandez2022watermarking}} & \textbf{TIACam} \\ 
\hline
Identity & 1.00 & 1.00 & \textbf{1.00} \\ 
Rotation (25°) & 0.27 & 1.00 & \textbf{1.00} \\ 
Crop (0.5) & 1.00 & \textbf{1.00} & 0.98 \\
Crop (0.1) & 0.02 & 0.98 & \textbf{1.00} \\ 
Resize (0.7) & 1.00 & 1.00 & \textbf{1.00} \\ 
Blur (2.0) & 0.25 & 1.00 & \textbf{1.00} \\
JPEG (50) & 0.96 & 0.97 & \textbf{0.99} \\ 
Brightness (2.0) & \textbf{0.99} & 0.96 & 0.97 \\ 
Contrast (2.0) & 1.00 & 1.00 & \textbf{1.00} \\
Hue (0.25) & 1.00 & 1.00 & \textbf{1.00} \\ 
Screenshot & 0.86 & 0.97 & \textbf{0.99} \\ 
\hline
\end{tabular}}
\vspace{-1.0em}
\end{table}

We next compare watermarking systems under geometric transformations in Table~\ref{tab:geom_compare}. Non-camera-oriented digital watermarking approaches (DWSF, MuST, WOFA) experience substantial performance drops as geometric severity increases, particularly under large rotations and scale changes. \mn\ maintains high accuracy across all geometric settings, demonstrating strong invariance to spatial misalignment and confirming the stability of the learned invariant feature space.

\begin{table}[h]
\centering
\caption{Bit accuracy (\%) under geometric distortions.}
\vspace{-0.75em}
\label{tab:geom_compare}
\resizebox{0.9\linewidth}{!}{
\begin{tabular}{lcccc}
\toprule
\textbf{Distortion} & \textbf{DWSF} \cite{guo2023practical} & \textbf{MuST} \cite{wang2024must} & \textbf{WOFA} \cite{liu2025watermarking} & \textbf{TIACam} \\
\midrule
Translation 10\% & 52.97 & 50.00 & 91.97 & \textbf{93.9} \\
Translation 25\% & 49.87 & 49.98 & \textbf{93.25} & 92.1 \\
Translation 50\% & 49.92 & 49.74 & 87.93 & \textbf{90.3} \\
\midrule
Rotation 15° & 50.21 & 49.79 & 95.26 & \textbf{99.2} \\
Rotation 30° & 49.74 & 49.73 & 94.24 & \textbf{97.8} \\
Rotation 45° & 49.60 & 49.82 & 90.63 & \textbf{95.5} \\
\midrule
Scaling $\pm$10\% & 53.30 & 49.98 & 95.72 & \textbf{97.4} \\
Scaling $\pm$20\% & 51.78 & 49.99 & 95.50 & \textbf{96.1} \\
Scaling $\pm$25\% & 51.08 & 50.00 & \textbf{95.02} & 93.4 \\
\bottomrule
\end{tabular}}
\end{table}

Finally, Table~\ref{tab:wm_extra_comparison} evaluates insertion-based watermarking methods (DWSF, WOFA, MuST) under real camera distortions. Although these systems are effective under purely digital perturbations, they degrade sharply in real capture scenarios. \mn\ consistently achieves higher extraction accuracy across bit lengths and distortion types, further demonstrating its strong generalization to physical-world artifacts.

\begin{table}[ht]
\vspace{-0.5em}
\centering
\caption{Bit accuracy (\%) of DWSF, WOFA, MuST, and \mn\ under screen camera, print camera, and screenshot distortions (30-bit and 100-bit).}
\vspace{-1.0em}
\resizebox{\columnwidth}{!}{
\begin{tabular}{lcccccc}
\toprule
\multirow{2}{*}{\textbf{Method}} 
& \multicolumn{2}{c}{\textbf{Screen Camera}} 
& \multicolumn{2}{c}{\textbf{Print Camera}} 
& \multicolumn{2}{c}{\textbf{Screenshots}} \\
\cmidrule(lr){2-3} \cmidrule(lr){4-5} \cmidrule(lr){6-7}
& 30 bits & 100 bits & 30 bits & 100 bits & 30 bits & 100 bits \\
\midrule
DWSF      & 77.4\% & 71.8\% & 69.4\% & 65.1\% & 64.9\% & 62.8\% \\
MuST  & 81.8\% & 80.6\% & 74.9\% & 72.3\% & 65.3\% & 64.1\% \\
WOFA       & 86.7\% & 81.1\% & 71.4\% & 70.2\% & 67.7\% & 64.8\% \\
\textbf{TIACam} & \textbf{99.1\%} & \textbf{98.2\%} & \textbf{96.6\%} & \textbf{95.1\%} & \textbf{97.4\%} & \textbf{95.2\%} \\
\bottomrule
\end{tabular}}
\label{tab:wm_extra_comparison}
\vspace{-1.5em}
\end{table}

Overall, these supplementary evaluations demonstrate that \mn\ provides state-of-the-art robustness not only in camera-based watermarking but also across a broad spectrum of digital and geometric distortions, confirming the general effectiveness of its distortion-invariant feature learning framework.

\section*{S5. Architectural Details}

\noindent \textbf{Architecture of the Auto-Augmentation Modules.} 
\mn auto-augmentor contains six learnable distortion modules: Additive, Photometric, Filtering, JPEG, Moiré, and Perspective, that jointly approximate realistic camera degradations. The Additive and Photometric modules share the same architecture: a 512-dimensional latent vector is expanded into a $8\times8\times128$ tensor through a fully-connected layer, followed by four ConvTranspose2d upsampling blocks that generate a full-resolution residual map using ReLU activations and a final Tanh layer. The Filtering module uses a 3-layer MLP (hidden size 256) to predict a $k\times k$ point-spread function ($k=3$), enforced to be non-negative via softplus and normalized to sum to one, then applied via depth-wise convolution. Both JPEG and Moiré modules adopt a U-Net with four encoder levels (64, 128, 256, 512 channels) and a 1024-channel bottleneck, paired with symmetric transposed-convolution decoders and skip connections; each predicts a residual that is added to the input image, enabling artifact-specific learning (compression ringing for JPEG, aliasing patterns for Moiré). Finally, the Perspective module uses a ResNet-18 encoder (with the classification layer removed), followed by a 3-layer MLP (256~$\rightarrow$~128~$\rightarrow$~9 units) that regresses a $3\times3$ homography matrix to simulate geometric warping. Together, these six differentiable modules capture noise, blur, photometric variation, JPEG compression, moiré interference, and perspective distortions observed in real camera pipelines.

\noindent \textbf{Architecture of the Feature Extractor and Discriminator.} 
\mn uses a lightweight MLP-based feature extractor together with a transformer-based discriminator to learn invariant alignment between image–text representations. The feature extractor takes a 768-dimensional CLIP embedding and processes it through an initial Linear–BatchNorm–ReLU–Dropout layer that expands the representation to 1024 units. This is followed by three ResidualBlocks, each containing two Linear layers with BatchNorm1d, ReLU activation, dropout (0.1), and a skip connection. A fusion block (Linear–BN–ReLU–Dropout) refines the intermediate representation, after which a projection head first reduces the dimensionality to 512 (Linear–BN–ReLU–Dropout) and then maps it to the final 1024-dimensional invariant space through a Linear–BatchNorm layer. Optional $\ell_2$ normalization is applied at the output. Overall, the feature extractor forms a 6-layer residual MLP designed to stabilize training and produce distortion-invariant embeddings.

The discriminator jointly processes the extracted image and text features to classify whether the pair originates from a real or adversarially generated match. Both embeddings are projected into a shared 512-dimensional hidden space via a Linear layer. A learnable \texttt{[CLS]} token is prepended, forming a 3-token sequence: \texttt{[CLS]}, image token, and text token. This sequence is passed through four stacked TransformerBlocks, each consisting of LayerNorm, an 8-head MultiheadAttention module, and a feed-forward MLP with GELU activation and dropout, with residual connections after both attention and MLP sub-layers. A final LayerNorm is applied to the output sequence, and the representation of the \texttt{[CLS]} token is fed into a fully connected layer to produce the 2-way classification logits. This combined architecture enables robust adversarial learning by aligning invariant features through the extractor while enforcing semantic consistency through the discriminator.

\section*{S6. Detailed Linear Probe Results}

Table~\ref{tab:linear_eval} provides the full numerical results corresponding to the linear probe comparisons summarized in Fig.~4 of the main paper. We report Top-1 and Top-5 accuracy for SimCLR, BYOL, Barlow Twins, VICReg, VIbCReg, and \mn\ across four datasets: CIFAR-100, Imagenette, MSCOCO, and Caltech-256, under six distortion types drawn from our camera-style pipeline (additive noise, photometric shift, perspective warp, JPEG compression, Moiré interference, and filtering noise).

These expanded results confirm the trend observed in the main paper: \mn\ consistently achieves the highest linear separability across all datasets and distortion categories. In particular, \mn\ shows large gains under geometric and appearance-heavy distortions, demonstrating that the learned invariant features retain strong semantic structure even under severe perturbations.

\begin{table*}[htb!]
\centering
\caption{Linear evaluation (Top-1 / Top-5 accuracy) on CIFAR-100, Imagenette, MSCOCO, and Caltech-256 under six distortion types.}
\vspace{-0.75em}
\small
\resizebox{\textwidth}{!}{
\begin{tabular}{llcccccccccccc}
\toprule
\multirow{2}{*}{\textbf{Dataset}} & \multirow{2}{*}{\textbf{Method}} 
& \multicolumn{2}{c}{\textbf{Additive}} 
& \multicolumn{2}{c}{\textbf{Photometric}} 
& \multicolumn{2}{c}{\textbf{Perspective}} 
& \multicolumn{2}{c}{\textbf{JPEG}} 
& \multicolumn{2}{c}{\textbf{Moir\'e}} 
& \multicolumn{2}{c}{\textbf{Filtering}} \\
\cmidrule(lr){3-4} \cmidrule(lr){5-6} \cmidrule(lr){7-8} 
\cmidrule(lr){9-10} \cmidrule(lr){11-12} \cmidrule(lr){13-14}
& & Top-1 & Top-5 & Top-1 & Top-5 & Top-1 & Top-5 & Top-1 & Top-5 & Top-1 & Top-5 & Top-1 & Top-5 \\
\midrule

\multirow{6}{*}{\textbf{CIFAR-100}} 
& SimCLR       & 72.1 & 83.3 & 71.0 & 79.1 & 70.5 & 86.7 & 62.4 & 76.5 & 70.2 & 81.9 & 71.3 & 86.8 \\
& BYOL         & 75.5 & 89.2 & 73.4 & 88.1 & 72.6 & 87.3 & 61.7 & 78.0 & 74.0 & 86.2 & 73.9 & 87.1 \\
& Barlow Twins & 71.8 & 88.7 & 69.6 & 85.7 & 72.9 & 86.9 & 65.8 & 77.6 & 75.5 & 87.0 & 78.2 & 86.6 \\
& VICReg       & 74.2 & 84.5 & 70.9 & 90.4 & 73.9 & 87.4 & 65.1 & 78.6 & 71.3 & 83.7 & 78.1 & 87.5 \\
& VIbCReg       & 72.9 & 88.5 & 72.9 & 88.4 & 76.9 & 87.4 & 66.1 & 82.6 & 76.3 & 81.7 & 77.1 & 87.5 \\
& \textbf{TIACam} & \textbf{81.5} & \textbf{95.1} & \textbf{77.7} & \textbf{94.2} & \textbf{83.8} & \textbf{93.6} & \textbf{73.2} & \textbf{85.0} & \textbf{83.9} & \textbf{89.8} & \textbf{82.6} & \textbf{93.4} \\
\midrule

\multirow{6}{*}{\textbf{Imagenette}}
& SimCLR       & 65.3 & 80.0 & 73.1 & 83.2 & 72.2 & 87.5 & 73.0 & 88.1 & 71.7 & 86.8 & 72.5 & 87.3 \\
& BYOL         & 72.2 & 86.1 & 72.0 & 83.0 & 73.0 & 88.2 & 73.8 & 88.9 & 72.4 & 87.1 & 73.2 & 87.9 \\
& Barlow Twins & 72.7 & 87.5 & 74.6 & 86.6 & 72.5 & 87.8 & 73.4 & 88.3 & 72.0 & 87.0 & 72.8 & 87.5 \\
& VICReg       & 75.6 & 90.3 & 74.4 & 87.4 & 73.4 & 88.5 & 74.1 & 89.0 & 72.7 & 87.4 & 73.5 & 88.1 \\
& VIbCReg       & 74.3 & 85.1 & 72.6 & 84.4 & 76.9 & 87.4 & 66.1 & 88.6 & 76.3 & 81.7 & 74.7 & 84.5 \\
& \textbf{TIACam} & \textbf{82.0} & \textbf{95.4} & \textbf{81.2} & \textbf{94.6} & \textbf{80.3} & \textbf{93.9} & \textbf{81.0} & \textbf{94.2} & \textbf{79.8} & \textbf{93.1} & \textbf{80.5} & \textbf{93.6} \\
\midrule

\multirow{6}{*}{\textbf{MSCOCO}}
& SimCLR       & 71.5 & 87.6 & 70.7 & 86.9 & 70.2 & 86.3 & 71.0 & 84.1 & 77.0 & 81.8 & 70.8 & 86.5 \\
& BYOL         & 72.3 & 88.2 & 71.4 & 87.5 & 70.8 & 86.8 & 71.6 & 87.6 & 79.5 & 86.2 & 71.3 & 86.9 \\
& Barlow Twins & 71.9 & 87.9 & 71.1 & 87.2 & 70.5 & 86.6 & 74.3 & 83.4 & 75.2 & 86.0 & 71.0 & 86.7 \\
& VICReg       & 74.7 & 88.5 & 71.7 & 87.8 & 71.0 & 87.1 & 75.9 & 85.9 & 78.7 & 83.5 & 71.5 & 87.2 \\
& VIbCReg       & 71.7 & 82.1 & 76.1 & 87.7 & 69.9 & 82.2 & 74.3 & 84.8 & 77.2 & 83.7 & 73.8 & 84.7 \\
& \textbf{TIACam} & \textbf{80.8} & \textbf{94.3} & \textbf{79.9} & \textbf{93.6} & \textbf{79.0} & \textbf{92.8} & \textbf{79.6} & \textbf{93.2} & \textbf{81.4} & \textbf{92.0} & \textbf{79.1} & \textbf{92.6} \\
\midrule

\multirow{5}{*}{\textbf{Caltech-256}}
& SimCLR       & 72.0 & 87.2 & 74.2 & 86.5 & 70.0 & 83.0 & 70.8 & 86.7 & 69.8 & 85.6 & 70.5 & 86.3 \\
& BYOL         & 71.8 & 89.8 & 78.0 & 87.1 & 73.6 & 86.5 & 71.4 & 88.3 & 70.2 & 86.0 & 71.0 & 86.6 \\
& Barlow Twins & 70.5 & 86.5 & 77.7 & 86.9 & 70.3 & 87.3 & 71.1 & 85.1 & 68.9 & 85.8 & 70.7 & 86.5 \\
& VICReg       & 72.2 & 90.1 & 73.4 & 87.5 & 72.9 & 83.8 & 71.7 & 87.6 & 70.5 & 86.2 & 71.2 & 86.9 \\
& VIbCReg       & 73.7 & 87.7 & 74.1 & 85.1 & 73.9 & 87.4 & 73.3 & 87.5 & 68.2 & 83.7 & 72.5 & 84.5 \\
& \textbf{TIACam} & \textbf{80.2} & \textbf{94.0} & \textbf{79.3} & \textbf{93.2} & \textbf{78.6} & \textbf{92.5} & \textbf{79.0} & \textbf{93.0} & \textbf{77.9} & \textbf{91.8} & \textbf{78.7} & \textbf{92.3} \\
\bottomrule
\end{tabular}
}
\label{tab:linear_eval}
\vspace{-1.0em}
\end{table*}

\section*{S7. Semantic Sensitivity under Invariance}

\begin{figure}[ht]
\vspace{-0.75em}
\centering
\includegraphics[width=0.85\linewidth]{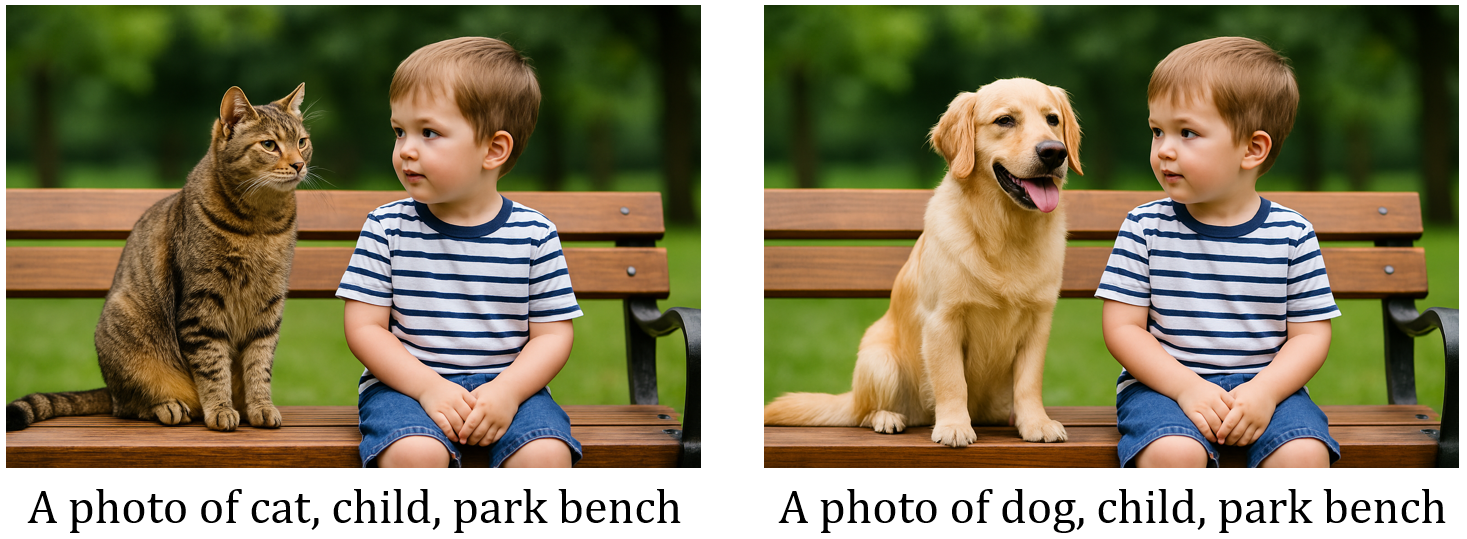}
\vspace{-0.5em}
\caption{Semantic disambiguation using controlled captions. Although the two images are visually similar, the model distinguishes them through small caption differences (e.g., “cat’’ vs.\ “dog’’), yielding higher similarity for matching image–text pairs and lower similarity for mismatched ones.}
\label{fig:cat_dogl}
\vspace{-1.0em}
\end{figure}

This experiment complements the “same-caption” analysis in the main paper by examining additional scenario: two visually similar images that differ only in subtle semantic details of their captions. Among 200 such image-caption pairs (e.g., “a photo of cat, child, park bench’’ vs.\ “a photo of dog, child, park bench’’ in Fig.~\ref{fig:cat_dogl}), \mn\ maintains strong alignment for true image-text pairs, achieving an average cosine similarity of 0.91. In contrast, even small changes in the caption reduce similarity to 0.70 on average. 

This behavior confirms that \mn\ is not merely invariant to distortions but also sensitive to semantic cues provided by text anchors. The invariant feature space preserves visual robustness while retaining the ability to disambiguate fine-grained linguistic differences, ensuring that each image remains both uniquely represented and semantically grounded.

\section*{S8. Failure Cases and Limitations}

\mn\ is designed for content-preserving distortions: perturbations that may alter the visual appearance but do not change the underlying semantic meaning of the image, which is typically the case for real camera-based watermark extraction. However, when this assumption is violated, the invariant feature space may no longer be recoverable, leading to failures in watermark extraction.

\begin{figure}[ht]
\centering
\includegraphics[width=0.75\columnwidth]{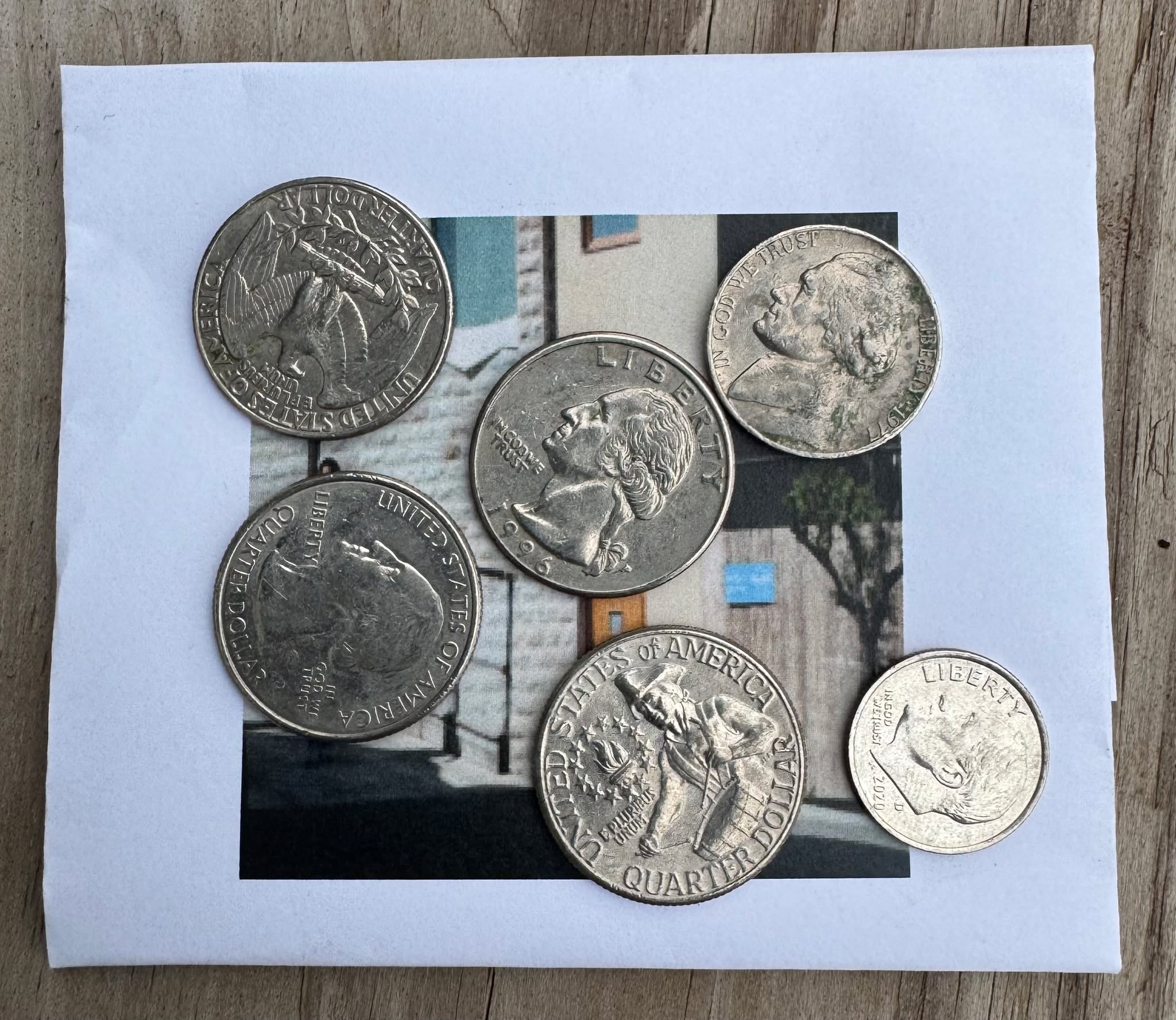}
\vspace{-0.7em}
\caption{Example failure case caused by severe occlusion.  
More than two-thirds of the image content is covered, removing or altering the semantic cues required by \mn\ to recover the invariant representation.  
Because the watermark is bound to the image’s underlying semantic meaning, a distortion that destroys or changes that meaning essentially yields a new image, making successful extraction impossible.}

\label{fig:fail_case}
\vspace{-1.0em}
\end{figure}

A representative failure mode occurs under severe occlusion or content removal. When a substantial portion of the image is blocked (empirically, more than two-thirds of the content; see Fig.~\ref{fig:fail_case}), the semantic cues required by the text anchor are no longer visible. In such cases, the feature extractor receives insufficient information to recover the intended invariant representation, causing the extracted features to drift away from the original manifold and leading to reduced or failed watermark decoding.

This limitation is consistent with the design of \mn: the watermark is associated with the underlying semantics of the image rather than its pixel-level appearance. When those semantics are largely removed or destroyed, the model can no longer align the extracted feature with the registered invariant signature. 
As a result, \mn\ remains robust under content-preserving distortions but cannot recover once the semantic content itself is lost or altered. In such cases, the distortion essentially produces a different new image in terms of meaning, and the model cannot align it with the original invariant signature.

\end{document}